\documentclass[twocolumn,prd,amsfonts,showpacs,epsfig,floats,nofootinbib,superscriptaddress,floatfix]{revtex4}
\usepackage{graphicx}
\usepackage{dcolumn}
\usepackage{bm}
\usepackage{longtable}
\usepackage{amsmath}

\begin{document}

\title{Cosmo-dynamics and dark energy with non-linear equation of state: a quadratic model}


\author{ Kishore N. Ananda}
\affiliation{Institute of Cosmology and Gravitation, University
of Portsmouth, Mercantile House, Portsmouth PO1 2EG, Britain}
\author{Marco Bruni}
\affiliation{Institute of Cosmology and Gravitation, University
of Portsmouth, Mercantile House, Portsmouth PO1 2EG, Britain}
\affiliation{Dipartimento di Fisica, Universit\`a di Roma ``Tor Vergata'',
via della Ricerca Scientifica 1, 00133 Roma, Italy}
\date{\today}


\begin{abstract}
We investigate the general relativistic dynamics of
Robertson-Walker models with a non-linear equation of state (EoS),
focusing on  the  quadratic case $P=P_0+\alpha \rho +\beta
\rho^2$. This may be taken to represent the Taylor expansion of
any arbitrary barotropic EoS, $P(\rho)$. With the right
combination of $P_0$, $\alpha$ and $\beta$, it  serves as a simple
phenomenological model for dark energy, or even unified dark
matter. Indeed we show that this simple model for the EoS can
produce a large variety of qualitatively different dynamical
behaviors that we classify using dynamical systems theory. An
almost universal feature is that accelerated expansion phases are
mostly natural for these non-linear EoS's. These are often
asymptotically de~Sitter thanks to the appearance of an {\it
effective cosmological constant}. Other interesting possibilities
that arise from the quadratic EoS  are closed models that can
oscillate with no singularity, models that bounce between infinite
contraction/expansion and models which evolve from a phantom
phase, asymptotically approaching a de Sitter phase instead of
evolving to a ``Big Rip''. In a second paper we investigate the
effects of the quadratic EoS in inhomogeneous and anisotropic
models, focusing in particular on singularities.

\end{abstract}

\pacs{98.80.Jk,~98.80.-k,~95.35.+d,~95.36.+x}

\maketitle

\section{Introduction}

Model building in cosmology requires two main ingredients: a
theory of gravity and a description of the matter content of the
universe.  In general relativity (GR) the gravity sector of the
theory is completely fixed, there are no free parameters. The
matter sector is represented in the field equations by the
energy-momentum tensor, and for a fluid the further specification of
an equation of state (EoS) is required. Apart from scalar fields,
typical cosmological fluids such as radiation or cold dark matter (CDM) are
represented by a {\it linear} EoS, $P={\it w}\rho $.

The combination of cosmic microwave background radiation
(CMBR)~\cite{CMBR,spergel}, large scale structure (LSS)~\cite{LSS}
and supernova type Ia (SNIa)~\cite{SNI} observations provides
support for a flat universe presently dominated by a component,
dubbed in general ``dark energy'',  causing an accelerated
expansion. The simplest form of dark energy is an {\it ad hoc}
cosmological constant $\Lambda$ term in the field equations, what
Einstein called his ``biggest blunder''. However,  although  the
standard  $\Lambda$CDM ``concordance'' model provides a rather
robust framework for the interpretation of present observations
(see e.g.~\cite{spergel,turner}), it requires a $\Lambda$ term
that is at odds by many order of magnitudes with theoretical
predictions~\cite{weinberg}. This has prompted theorists to
explore possible dark energy sources for the acceleration that go
beyond the standard but unsatisfactory $\Lambda$. With various
motivations, many authors have attempted to describe dark energy
as quintessence, {\it k}-essence or a ghost field, i.e. with
scalar fields with various properties. There have also been
attempts to describe dark energy by a fluid with a specific
non-linear EoS like the Chaplygin gas~\cite{KMP}, generalized
Chaplygin gas~\cite{GCG}, van der Waals fluid~\cite{GK}, wet dark
fluid~\cite{HN} and other specific gas EoS's~\cite{CTTC}.
 Recently, various ``phantom models'' (${\it
w}=P/\rho<-1 $) have also been considered~\cite{RRC, sahni}. More
simply, but also with a higher degree of generality, many authors
have focused on phenomenological models where dark energy is
parameterized by assuming a $w=P/\rho=w(a)$, where $a=a(t)$ is the
expansion scale factor (see e.g.~\cite{bruce,will}).

Another possibility is to advocate a modified theory of gravity.
At high energies, modification of gravity beyond general
relativity could come from extra dimensions, as required in string
theory. In the brane world~\cite{SMS,BMW,DL,RM} scenario the extra
dimensions produce a term quadratic in the energy density in the
effective 4-dimensional energy-momentum tensor. Under the
reasonable assumption of neglecting 5-dimensional Weyl tensor
contributions on the brane, this quadratic term has the very
interesting effect of suppressing anisotropy at early enough
times. In the case of a Bianchi I brane-world cosmology containing
a scalar field with a large kinetic term the initial expansion is
quasi-isotropic~\cite{MSS}. Under the same assumptions, Bianchi I
and Bianchi V brane-world cosmological models containing standard
cosmological fluids with linear EoS  also behave in a similar
fashion\footnote{This only requires $P/ \rho={\it w}>0 $, as
opposed to ${\it w}>1 $ in the GR case. In the case of
ekpyrotic/cyclic and pre-big bang models the initial expansion is
only isotropic if ${\it w}>1 $ as in the case of GR
~\cite{EWST}.}~\cite{CS}, and the same remains true for more
general homogeneous models~\cite{coley1,coley2} and even some
inhomogeneous exact solutions~\cite{coley3}. Finally, within the
limitations of a perturbative treatment, the
quadratic-term-dominated isotropic brane-world models have been
shown to be local past attractors in the larger phase space of
inhomogeneous and anisotropic models~\cite{DGBC, GDCB}. More
precisely, again assuming that the 5-d Weyl tensor contribution to
the brane can be neglected, perturbations of the isotropic models
decay in the past. Thus in the brane scenario the observed high
isotropy of the universe is the natural outcome of {\it generic
initial conditions}, unlike in GR where in general cosmological
models with a standard energy momentum tensor are highly
anisotropic in the past (see e.g. \cite{LL}).

Recently it has been shown that loop quantum gravity
corrections result in a modified Friedmann equation~\cite{KV},
with the modification appearing as a negative term which is
quadratic in the energy density. Further motivation for
considering a quadratic equation of state comes from recent
studies of {\it k}-essence fields as unified dark matter (UDM)
models\footnote{These attempt to provide a unified model for both
the dark matter and the dark energy components necessary to make
sense of observations.}~\cite{GH,RS}. The general {\it k}-essence
field can be described by a fluid with a closed-form barotropic
equation of state. The UDM fluid discussed in~\cite{GH} has a
non-linear EoS of the form $P\propto \rho^2$ at late times. More
recently, it has been shown that any purely kinetic {\it
k}-essence field can be interpreted as an isentropic  perfect
fluid with an EoS of the form $P=P(\rho)$~\cite{DTF}. Also, low
energy dynamics of the Higgs phase for gravity have been shown to
be equivalent to the irrotational flow of a perfect fluid with
equation of state $P=\rho^2$~\cite{ACLMW}.

Given the isotropizing effect that the quadratic energy density
term has at early times in the brane scenario this then prompts
the question: can a term quadratic in the energy density have the
same effect in general relativity. This question is non-trivial as
the form of the equations in the two cases is quite different. On
the brane,  for a given EoS the effective 4-dimensional Friedmann
and Raychaudhuri equations are modified, while the continuity
equation is identical to that of  GR. With the introduction of a
quadratic EoS in GR, the Friedman equation remains the same, while
the continuity and Raychaudhuri equations are
modified\footnote{With Respect to the case of the same EoS with
vanishing quadratic term.}.

Taking into account this question (to be explored in detail in
Paper II~\cite{AB}), the diverse motivations  for a quadratic
energy density term mentioned above and with the dark energy
problem in mind, in this paper we explore the  GR dynamics of
homogeneous isotropic Robertson-Walker models with a quadratic
EoS, $P=P_0+\alpha \rho +\beta \rho^2$. This is the simplest model
we can consider without making any more specific assumptions on
the EoS~\cite{MV}. It represents the first terms of the Taylor
expansion of {\it  any} EoS function $P=P(\rho)$ about $\rho=0$.
It can also be taken to represent (after re-grouping of terms) the
Taylor expansion about the present energy density $\rho_0$, see
\cite{MV}. In this sense therefore the out-coming dynamics is very
general. Indeed it turns out  that this simple model can produce a
large variety of qualitatively different dynamical behaviors that
we classify using dynamical systems theory~\cite{WE, AP}. An
outcome of our analysis is that accelerated expansion phases are
mostly natural for non-linear EoS's. These are {\it in general}
asymptotically de Sitter thanks to the appearance of an {\it
effective cosmological constant}. This suggests that an EoS with
the right combination of $P_0$, $\alpha$ and $\beta$ may provide a
good and simple phenomenological model for  UDM, or at least for a
dark energy component. Other interesting possibilities that arise
from the quadratic EoS  are closed models that can oscillate with
no singularity, models that bounce between infinite
contraction/expansion and models which evolve from a phantom
phase, asymptotically approaching a de Sitter phase instead of
evolving to a ``big rip'' or other pathological future states
\cite{RRC,BLJM,NOT}.

As mentioned before, the question of the dynamical effects  the
quadratic energy density term  has on the anisotropy  in GR is
explored in Paper II~\cite{AB}. There we analyze Bianchi I and V
models with the EoS $P=\alpha\rho +\beta \rho^2$, as well as
perturbations of the isotropic past attractor of those  models
that are singular in the past. We anticipate that Bianchi I and V
non-phantom models with $\beta>0$ have an isotropic singularity,
i.e. they are asymptotic in the past to a certain isotropic model,
and that perturbations of this model decay in the past. Phantom
anisotropic models with $\beta>0$ are necessarily asymptotically
de~Sitter in the future, but the shear anisotropy  dominates in
the past. For $\beta<0$ all models are anisotropic in the past,
while their specific future evolution depends on the value of
$\alpha$.

The paper is organized as follows. In section~\ref{sec2} we
outline the setup and the three main cases we will investigate. In
section~\ref{sec3}, we study the dynamics of isotropic
cosmological models in the high energy limit (neglecting the $P_0$
term). We find the critical points, their stability nature and the
occurrence of bifurcations of the dynamical system. In
section~\ref{sec4}, we consider the low energy limit (neglecting
the $\rho^2$ term). The full system is then analyzed in
section~\ref{sec5}, showing the qualitatively different behavior
with respect to the previous cases. We then finish with some
concluding remarks and an outline of work in progress in
section~\ref{sec6}. Units are such that $8\pi G/c^4=1$.

\section{Cosmology with a quadratic EoS}\label{sec2}

\subsection{Dynamics with non-linear EoS}

The evolution of Robertson-Walker isotropic models with no
cosmological constant $\Lambda$ term is given in GR by the
following non-linear planar autonomous dynamical system:
\begin{eqnarray}
\dot{\rho}&=& -3 H \left( \rho + P \right), \label{energycons}\\
\dot{H} &=& -H^2 - \frac{1}{6}\left( \rho + 3P \right),\label{Ray}
\end{eqnarray}
where $H$ is the Hubble expansion function, related to the scale
factor $a$ by $H=\dot{a}/a$. In order to close this system of
equations, an EoS must be specified, relating
the isotropic pressure $P$ and the energy density $\rho$. When an
EoS $P=P(\rho)$ is given, the above system admits a first
integral, the Friedman equation
\begin{equation}\label{Friedman}
H^{2} = \frac{1}{3}\rho - \frac{K}{a^2},
\end{equation}
\noindent where $K$ is the curvature, $K=0,\pm 1$ as usual for
flat, closed and open models.

Here we are interested in exploring the general dynamical features
of a non-linear EoS $P=P(\rho)$. Before considering the specific
case of a quadratic EoS, we note some important general points.

First, it is immediately clear from Eq.~(\ref{energycons}) that an
effective cosmological constant is achieved whenever there is an
energy density value $\rho_{\Lambda}$ such that $P(\rho_{\Lambda})
= -\rho_{\Lambda}$.
More specifically:\\

\noindent {\bf Remark 1.} If for a given EoS function $P=P(\rho)$
there exists a $\rho_\Lambda$ such that $P(\rho_\Lambda)=
-\rho_\Lambda$, then $\rho_\Lambda$ has the dynamical role of an
effective cosmological constant.\\

\noindent {\bf Remark 2.} A given EoS $P(\rho)$ may admit more
than one point $\rho_\Lambda$. If these points exist,  they  are
fixed points of Eq. (\ref{energycons}).\\

\noindent {\bf Remark 3.} From  Eq.~(\ref{Ray}), since
$\dot{H}+H^2 = \ddot{a}/a$, an accelerated phase is achieved
whenever $P(\rho) < -\rho/3$.\\

\noindent {\bf Remark 4.} Remark 3 is only valid in GR, and a
different condition will be valid in other theories of gravity.
Remarks 1 and 2, however, are only based on conservation of
energy, Eq.~(\ref{energycons}). The latter is also valid (locally)
in inhomogeneous models, provided that the time derivative is
taken to represent the derivative along the fluid flow lines (e.g.
see~\cite{ellis_varenna71}), and is a direct consequence of
$T^{ab}{}_{;b}=0$. Thus Remarks 1 and 2 are valid in any gravity
theory where $T^{ab}{}_{;b}=0$, as well as (locally) in inhomogeneous models.\\

Second, assuming expansion, $H>0$, we may rewrite
Eq.~(\ref{energycons}) as:
\begin{equation}
\frac{d\rho}{d\tau}=-3\left[\rho+P(\rho)\right], \label{encon1}
\end{equation}
where $\tau=\ln a$.  Eq.~(\ref{encon1}) is a 1-dimensional
dynamical system with fixed point(s) $\rho_\Lambda$(s), if they
exist. If $\rho +P(\rho)<0$ the fluid violates the null energy
condition~\cite{carroll,visser}  and Eq. (\ref{energycons}) implies what
has been dubbed phantom  behavior~\cite{RRC} (cf.~\cite{LM}),
i.e.\ the fluid behaves counter intuitively in that the energy
density increases (decreases) in the future for an expanding
(contracting) universe.
Then:\\

\noindent {\bf Remark 5.} Any point $\rho_\Lambda$ is an attractor
(repeller) of the evolution during expansion (the autonomous system (\ref{encon1})) if
$\rho+P(\rho)<0$ ($>0$) for $\rho<\rho_\Lambda$ and
$\rho+P(\rho)>0$ ($<0$) for $\rho>\rho_\Lambda$.\\

\noindent {\bf Remark 6.} Any point $\rho_\Lambda$ is a
shunt\footnote{This is a fixed point which is  an attractor for one
side of the phase line and a repeller for the other~\cite{AP}.}
 of the autonomous system Eq.~(\ref{encon1}) if either
$\rho+P(\rho)<0$ on both sides of  $\rho_\Lambda$,  or
$\rho+P(\rho)>0$ on both sides of $\rho>\rho_\Lambda$. In this
case the fluid is respectively phantom or standard on both
sides.\\

Let's now consider the specific case of a general quadratic EoS of
the form:
\begin{equation}\label{QuadEoS}
P = {P}_{o} + \alpha\rho + \beta{{\rho}^2}.
\end{equation}

\noindent The parameter $\beta$ sets the characteristic energy
scale $\rho_{c}$ of the quadratic term as well as it's sign
$\epsilon$
\begin{equation}
\beta=\frac{\epsilon}{{\rho}_{c}}.
\end{equation}

\noindent {\bf Remark 7.} Eq. (\ref{QuadEoS}) represents the
Taylor expansion, up to ${\cal O}(3)$, of {\it any} barotropic EoS
function $P=P(\rho)$ about $\rho=0$. It also represents, after
re-grouping of terms, the Taylor expansion about the present
energy density value $\rho_0$~\cite{MV}. In this sense, the
dynamical system (\ref{energycons},\ref{Ray}) with (\ref{QuadEoS})
is {\it general}, i.e. it represents the late evolution, in GR,
of {\it any} cosmological model with non-linear barotropic EoS
approximated by Eq. (\ref{QuadEoS}).\\

The usual scenario for a cosmological fluid is a
standard linear EoS ($P_0=\beta=0$), in which case $\alpha=w$ is
usually restricted to the range $-1<\alpha<1$. For the sake of
generality, we will consider values of $\alpha$ outside this
range, considering dynamics only restricted by the  request
that $\rho\geq 0$. The first term in Eq~(\ref{QuadEoS}) is a
constant pressure term which  in general  becomes important in what we call
the low energy regime. The second term is the standard linear term
usually considered, with
\begin{equation}
\alpha=\frac{dP}{d\rho}\Bigg|_{\rho=0}.
\end{equation}
If it is positive, $\alpha$ has an interpretation in terms of the
speed of sound of the fluid in the limit $\rho\rightarrow 0$,
$\alpha=c_s^2$. The third term is quadratic in the energy density
and will be important in what we call the high energy regime.

In the following, we first split the analysis of the dynamical
system Eqs. (\ref{energycons},~\ref{Ray},~\ref{QuadEoS}) in two
parts, the high energy regime where we neglect $P_0$ and the low
energy regime where we set $\beta=0$, then we consider the full
system with EoS (\ref{QuadEoS}).  Using only the energy
conservation Eq.~(\ref{energycons}) we list the various sub-cases,
also briefly anticipating  the main dynamical features coming out
of the analysis in Sections \ref{sec3}, \ref{sec4} and \ref{sec5}.

\subsection{Quadratic EoS for the high energy regime}

In the high energy regime we consider the restricted equation of
state:
\begin{equation}
P_{HE}= \alpha\rho + \frac{\epsilon {\rho}^2}{\rho_{c}}.
\end{equation}

\noindent The energy conservation Eq.~(\ref{energycons}) can be
integrated in general to give:
\begin{eqnarray} \label{rhohe}
\rho_{HE}(a) &=& \frac{A(\alpha+1)\rho_{c}}{a^{3(\alpha+1)} - \epsilon A},\\
A &=& \frac{\rho_{o}  a_{o}^{3(\alpha+1)}}{(\alpha+1)\rho_{c}
+\epsilon \rho_{o}},
\end{eqnarray}
where $\rho_o$, $a_o$ represent the energy density and scale
factor at an arbitrary time $t_o$. This is valid for all values of
$\epsilon$, $\rho_{c}$ and $\alpha$, except for $\alpha\neq-1$. In
the case $\alpha=-1$ the evolution of the energy density is:
\begin{eqnarray} \label{rhoheb}
\rho_{HE}(a) &=& \left[ \frac{1}{\rho_{o}}+
\frac{3\epsilon}{\rho_{c}} \ln \left( \frac{a}{a_{o}} \right)
\right]^{-1}.
\end{eqnarray}

\noindent The EoS with this particular choice of parameters has
already been considered as a possible  dark energy model
\cite{NOT, HS}. We will concentrate on the broader class of models
where $\alpha\neq-1$.

In Section \ref{sec3} we will give a dynamical system analysis of
the high energy regime, but it is first useful to gain some
insight directly from Eq.~(\ref{rhohe}).

We start by defining
\begin{equation}
\label{lambda}
\rho_\Lambda :=-\epsilon
(1+\alpha)\rho_c,
\end{equation}
 noticing that this is  an effective positive cosmological
constant point only if $\epsilon
(1+\alpha) <0$.  It is then convenient to rewrite
Eq.~(\ref{rhohe}) in three different ways, defining $a_{\star}=
|A|^{1/3(\alpha+1)}$, each representing two different
subcases.\\

\noindent {\bf A:} $ \epsilon (1+\alpha)>0$, $\rho_\Lambda<0$,
\begin{equation}
\rho=\frac{|1+\alpha|\rho_c}{\left(\frac{a}{a_\star}\right)^{3(1+\alpha)} -1}.
\label{rho1}
\end{equation}

\noindent {\bf A1:} $\epsilon>0$, $(1+\alpha)>0$. In this case
$a_\star<a<\infty$, with $\infty>\rho >0$. Further restrictions on
the actual range of values that $a$ and $\rho$ can take may come
from the geometry. For a subset of appropriate initial conditions
closed (positively curved) models may expand to a maximum $a$
(minimum $\rho$) and re-collapse, and for $\alpha<-1/3$ not all
closed models have a past singularity at $a=a_\star$, having
instead a bounce at a minimum $a$ (maximum $\rho$).\\

\noindent {\bf A2:} $\epsilon<0$, $(1+\alpha)<0$. In this case
$0<a<a_\star$, with $0 <\rho <\infty$, and the fluid
exhibits phantom behavior. All models have a future singularity at
$a=a_\star$, but in general closed models contract from a past
singularity, bounce at a minimum $a$ and $\rho$, then
re-expand to the future singularity (we will refer to this as a phantom bounce).\\

\noindent {\bf B:} $ \rho_\Lambda>0$,
$\rho>\rho_\Lambda$,
\begin{equation}
\rho=\frac{\rho_\Lambda}{1-\left(\frac{a}{a_\star}\right)^{3(1+\alpha)}}.
\label{rho2}
\end{equation}

\noindent {\bf B1:} $\epsilon>0$, $(1+\alpha)<0$, $A>0$. In this
case $a_\star<a<\infty$, with $\infty>\rho >\rho_\Lambda$. As in
case {\bf A1}, further restrictions on the actual range of values
that $a$ and $\rho$ can take may come from the geometry. For a
subset of initial conditions closed models may expand to a maximum
$a$ (minimum $\rho$) and re-collapse, while for another subset
closed models don't have a past singularity at $a=a_\star$,
having instead a bounce at a minimum $a$ (maximum $\rho$).\\

\noindent {\bf B2:} $\epsilon<0$, $(1+\alpha)>0$, $A<0$. In this
case $0<a<a_\star$, with $\rho_\Lambda<\rho <\infty$. As in the
case {\bf A2}, the fluid has a phantom  behavior. All models
have a future singularity at $a=a_\star$, with closed models
contracting  from a past singularity to a minimum $a$ and $\rho$
before re-expanding.\\

\noindent {\bf C:} $ \rho_\Lambda>0$,
$\rho<\rho_\Lambda$,
\begin{equation}
\rho=\frac{\rho_\Lambda}{1+\left(\frac{a}{a_\star}\right)^{3(1+\alpha)}}.
\label{rho3}
\end{equation}

\noindent {\bf C1:} $\epsilon>0$, $(1+\alpha)<0$, $A<0$. In this
case $0<a<\infty$, with $0<\rho <\rho_\Lambda$. The fluid behaves
in a phantom manner but avoids the future singularity and instead
evolves to a constant energy density $\rho_\Lambda$. Closed models,
however, typically bounce with a minimum $\rho$ at a finite $a$.\\

\noindent {\bf C2:} $\epsilon<0$, $(1+\alpha)>0$, $A>0$. In this
case $0<a<\infty$, with $\rho_\Lambda>\rho >0$. Again, closed
models may evolve within restricted ranges of $a$ and $\rho$, even
oscillating, for $\alpha\geq -1/3$, between maxima and minima of
$a$ and $\rho$.

\subsection{Low energy regime: affine EoS}

In the low energy regime we consider the affine equation of state:
\begin{equation}
P_{LE}= P_{o} + \alpha\rho.
\end{equation}

\noindent This particular EoS has been investigated as a possible
dark energy model~\cite{HN, Babi}, however, only spatially flat
Friedmann models where considered. The scale factor dependence of
the energy density  is:
\begin{eqnarray}\label{rhole}
\rho_{LE}(a)= -\frac{P_{o}}{(\alpha+1)} + B a^{-3(\alpha+1)},\\
B={\left[ \frac{P_{o}}{(1+\alpha)} + \rho_{o}
\right]}{a_{o}}^{3(1+\alpha)}.
\end{eqnarray}

\noindent  This is valid for all values of $P_{o}$ and $\alpha$
except $\alpha\neq-1$. In the case $\alpha=-1$, the evolution of
the energy density is:
\begin{eqnarray} \label{rholeb}
\rho_{LE}(a) &=&  {\rho_{o}} - 3{P_{o}} \ln \left( \frac{a}{a_{o}}
\right) ,
\end{eqnarray}

\noindent As in the high energy case, we will concentrate on the
broader class of models where $\alpha\neq-1$.

In Section \ref{sec4} we present the dynamical system analysis of
the low energy regime, but first let us gain some insight from Eq.
(\ref{rhole}). As with the high energy case, in many cases the fluid
violates the null energy condition ($\rho+P<0$) and exhibit
phantom behavior. Defining
\begin{equation}
\label{lambdatilde}
\tilde{\rho}_\Lambda :=-P_{o}/(1+\alpha),
\end{equation}
we see that a positive effective cosmological constant point
exists, $\tilde{\rho}_\Lambda > 0$, only if $P_{o}/(1+\alpha) <0$.
Eq.~(\ref{rhole}) can be rewritten in three different ways,
defining $\tilde{a}_{\star}= |B|^{1/3(\alpha+1)}$, each
representing two different subcases.\\

\noindent {\bf D:} $P_{o}/(1+\alpha)>0$, $\tilde{\rho}_\Lambda<0$,
\begin{equation}\label{rho4}
\rho=-\frac{P_{o}}{(\alpha+1)}+\left(\frac{a}{\tilde{a}_\star}
\right)^{-3(1+\alpha)}.
\end{equation}

\noindent {\bf D1:} $P_{o}>0$, $(1+\alpha)>0$. In this case
$0<a<\infty$, with $\infty>\rho >-|\tilde{\rho}_\Lambda|$. The
geometry places further restrictions on the values that $a$ and
$\rho$ can take. The subset of open models (negative curvature)
are all non-physical as they evolve to the $\rho<0$ region of the
phase space. The spatially flat models expand to a maximum $a$
(when $\rho=0$) and recollapse. The closed (positively curved)
models expand to a maximum $a$ (minimum $\rho$) and recollapse,
and for $-1\leq\alpha<-1/3$ a subset of closed models oscillate
between a maximum and minimum $a$ (minimum and maximum $\rho$).\\

\noindent {\bf D2:} $P_{o}<0$, $(1+\alpha)<0$. In this case
$0<a<\infty$, with $-|\tilde{\rho}_\Lambda| <\rho <\infty$. In
this case the fluid exhibits phantom behavior. The subset of open
models are all non-physical as they evolve from the $\rho<0$
region of the phase space. The spatially flat models contract,
bounce at a minimum $a$ when $\rho=0$ and re-expand in the future.
The closed models contract, bounce at a minimum $a$ and $\rho$,
then re-expand in the future.\\

\noindent {\bf E:} $\tilde{\rho}_\Lambda>0$,
$\rho>\tilde{\rho}_\Lambda$,
\begin{equation}\label{rho5}
\rho=\tilde{\rho}_\Lambda+\left(\frac{a}{\tilde{a}_\star}
\right)^{-3(1+\alpha)}.
\end{equation}

\noindent {\bf E1:} $P_{o}>0$, $(1+\alpha)<0$, $B>0$. In this case
$0<a<\infty$, with $\tilde{\rho}_\Lambda<\rho<\infty$. As in the
case {\bf D2}, the fluid behaves in a phantom manner. The flat and
open models are asymptotically de Sitter in the past, when their
energy density approaches a finite value ($\rho \rightarrow
\tilde{\rho}_\Lambda$ as $a\rightarrow0$), and when
$\tilde{\rho}_\Lambda $ becomes negligible in Eq.~(\ref{rho5})
they evolve as standard linear phantom models, reaching a future
singularity in a finite time  ($\rho \rightarrow \infty$ as
$a\rightarrow\infty$). The closed models contract to a
minimum $a$ (minimum $\rho$), bounce and re-expand.\\

\noindent {\bf E2:} $P_{o}<0$, $(1+\alpha)>0$, $B>0$. In this case
$0<a<\infty$, with $\infty>\rho>\tilde{\rho}_\Lambda$. All flat
and open models expand from a singularity and asymptotically
evolve to a de Sitter model, with $\rho=\tilde{\rho}_\Lambda$. The
closed models evolve from a contracting de Sitter model to minimum
$a$ (maximum $\rho$), bounce and then evolve to an expanding de Sitter model.\\

\noindent {\bf F:} $\tilde{\rho}_\Lambda>0$,
$\rho<\tilde{\rho}_\Lambda$,
\begin{equation}\label{rho6}
\rho=\tilde{\rho}_\Lambda-\left(\frac{a}{\tilde{a}_\star}
\right)^{-3(1+\alpha)}.
\end{equation}

\noindent {\bf F1:} $P_{o}>0$, $(1+\alpha)<0$, $B<0$.  In this
case $0<a<\infty$, with $\tilde{\rho}_\Lambda>\rho>-\infty$. The
subset of open models are all non-physical as they evolve to the
$\rho<0$ region of the phase space. The flat models evolve from an
expanding de Sitter phase to a contracting de Sitter phase. The
closed models oscillate between a maximum and minimum $a$ (minimum
and maximum $\rho$).\\

\noindent {\bf F2:} $P_{o}<0$, $(1+\alpha)>0$, $B<0$.  In this
case $0<a<\infty$, with $-\infty<\rho<\tilde{\rho}_\Lambda$. The
fluid exhibits phantom behavior. The open models are all
non-physical as they evolve from the $\rho<0$ region of the phase
space. The flat and closed models evolve from a contracting de
Sitter phase, bounce at  minimum $a$ and $\rho$,  then re-expand,
asymptotically approaching a expanding de Sitter phase.

\subsection{The full quadratic EoS}

In Section \ref{sec5} we present the dynamical system analysis of
the full quadratic EoS models given by Eq.~(\ref{QuadEoS}), but
again we first study the form of $\rho(a)$ implied by conservation
of energy, Eq. (\ref{energycons}). As with the previous cases the
fluid can violate the null energy condition ($\rho+P<0$) and
therefore may exhibit phantom behavior. The system may admit two
(possibly negative) effective cosmological constant points:
\begin{eqnarray}
\rho_{\Lambda,1} &:=& \frac{1}{2\beta}\left[-(\alpha+1) + \sqrt{\Delta} \right],\\
\rho_{\Lambda,2} &:=& \frac{1}{2\beta}\left[-(\alpha+1) - \sqrt{\Delta} \right],
\end{eqnarray}
if
\begin{equation}
\Delta := (\alpha+1)^2 - 4\beta P_{o}
\end{equation}
is non negative. Clearly, the existence of the effective
cosmological points depends on the values of the parameters in the
EoS. This in turn affects the functional form of $\rho(a)$. In
order to find $\rho(a)$ the following integral must be evaluated:
\begin{eqnarray}
-3\ln \left( \frac{a}{a_{o}} \right)= \int^{\rho}_{\rho_{o}}
\frac{d\rho}{{P}_{o} + (\alpha+1)\rho + \beta{{\rho}^2}}.
\end{eqnarray}

\noindent This is done  separately for the cases when
no effective cosmological points exist ($\Delta<0$), when one
cosmological point exist, $\rho_{\Lambda,1} = \rho_{\Lambda,2} =
\bar{\rho}_{\Lambda} \neq 0 $ ($\Delta=0$) and when two cosmological
points exist, $\rho_{\Lambda,1} \neq \rho_{\Lambda,2} \neq 0 $
($\Delta>0$). We now consider these three separate sub-cases.\\

\noindent {\bf G:} $(1+\alpha)^{2}<4\beta P_{o}$, $\Delta<0$,
\begin{eqnarray}
\rho &=& \frac{ \Gamma - \sqrt{|\Delta|}\tan \left( \frac{3}{2}
\sqrt{|\Delta|} \ln \left( \frac{a}{a_{o}} \right)\right) }{
2\beta + \frac{2\beta}{\sqrt{|\Delta|}}\Gamma \tan \left(
\frac{3}{2} \sqrt{|\Delta|} \ln \left( \frac{a}{a_{o}}
\right)\right) } - \frac{(\alpha+1)}{2\beta},\nonumber\\
\nonumber\\
\Gamma &=& 2 \beta \rho_{o} + (\alpha+1).\label{rho7}
\end{eqnarray}

\noindent {\bf G1:} $\beta>0$, $P_{o}>0$. In this case $a_{1}< a <
a_{2}$ (where $a_{1}<a_{2}$), with $\infty>\rho>-\infty$. The
fluid behaves in a standard manner and all models have a past
singularity at $a=a_{1}$. All open models are non-physical as they
evolve to the $\rho<0$ region of the phase space. The flat models
expand to a maximum $a$ ($\rho = 0$) and then re-collapse. The
closed models can behave in a similar manner to flat models except
they reach a minimum $\rho$ before re-collapsing. Some closed
models oscillate between maxima and minima $a$ and $\rho$.\\

\noindent {\bf G2:} $\beta<0$, $P_{o}<0$. In this case $a_{1}< a <
a_{2}$ (where $a_{1}<a_{2}$), with $-\infty<\rho<\infty$. The
fluid behaves in a phantom manner. All open models are
non-physical as they evolve from the $\rho<0$ region of the phase
space. The flat and closed models represent phantom bounce models,
that is they evolve from a singularity at $a=a_{1}$
($\rho=\infty$), contract to a minimum $a$ (minimum $\rho$) and
then re-expand to the future singularity at $a=a_{2}$.\\

\noindent {\bf H:} $(1+\alpha)^{2}=4\beta P_{o}$, $\Delta=0$,
\begin{eqnarray}
\rho &=& \bar{\rho}_{\Lambda} + \frac{1} { 3 \beta \ln \left(
\frac{a}{a_{o}} \right) + \frac{2\beta}{\Gamma}  } .\label{rho8}
\end{eqnarray}

\noindent {\bf H1:} $\beta>0$, $P_{o}>0$, $\rho<\bar{\rho}_{\Lambda}$.
In this case $0 < a < a_{1}$ with $\bar{\rho}_{\Lambda}>\rho>-\infty$.
The fluid behaves in a standard manner. The subset of open models
are all non-physical as they evolve to the $\rho<0$ region of the
phase space. The flat models evolve from an expanding de Sitter
phase to a contracting de Sitter phase. The closed models
oscillate between maxima and minima $a$ and $\rho$.\\

\noindent {\bf H2:} $\beta>0$, $P_{o}>0$,
$\rho>\bar{\rho}_{\Lambda}$. In this case $a_{1} < a < \infty $
with $\infty> \rho> \bar{\rho}_{\Lambda}$ and the fluid behaves in
a standard manner. If $\bar{\rho}_{\Lambda}>0$, the open and flat
models evolve from a past singularity ($a=a_{1}$) and evolve to a
expanding de Sitter phase. For a subset of initial conditions
closed models may expand to a maximum $a$ (minimum $\rho$) and
re-collapse, while for another subset closed models avoid a past
singularity, instead having a bounce at a minimum $a$ (maximum
$\rho$). If $\bar{\rho}_{\Lambda}<0$, the open models are
non-physical, while flat and closed models represent recollapse models.\\

\noindent {\bf H3:} $\beta<0$, $P_{o}<0$, $\rho<\bar{\rho}_{\Lambda}$.
In this case $a_{1}<a<\infty $ with $-\infty< \rho<
\bar{\rho}_{\Lambda}$. The fluid behaves in a phantom manner. The open
models are all non-physical as they evolve from the $\rho<0$
region of the phase space. The flat and closed models evolve from
a contracting de Sitter phase, bounce at minimum $a$ and $\rho$,
then re-expand, asymptotically approaching an expanding de Sitter
phase.\\

\noindent {\bf H4:} $\beta<0$, $P_{o}<0$,
$\rho>\bar{\rho}_{\Lambda}$. In this case $0 < a < a_{1} $ with
$\bar{\rho}_{\Lambda}<\rho<\infty$ and the fluid behaves in a
phantom manner. All models have a future singularity at $a=a_{1}$.
If $\bar{\rho}_{\Lambda}>0$, closed models contract from a past
singularity to a minimum $a$ and $\rho$ before re-expanding
(phantom bounce), while flat and open models are asymptotic to
generalized de Sitter models in the past. If
$\bar{\rho}_{\Lambda}<0$, open models are non-physical, while
flat and closed models contract from a past singularity to
a minimum $a$ and $\rho$ before re-expanding.\\

\noindent {\bf I:} $(1+\alpha)^{2}>4\beta P_{o}$, $\Delta>0$,
\begin{eqnarray}
\rho &=&\frac{\rho_{\Lambda,2}  \left(
\frac{a}{a_{o}} \right)^{-3\sqrt{\Delta}} -\rho_{\Lambda,1} C}{ \left(
\frac{a}{a_{o}} \right)^{-3\sqrt{\Delta}}-C} ,  \\
C&=&\frac{ \rho_o-\rho_{\Lambda,2}}{\rho_o-\rho_{\Lambda,1}}.
\end{eqnarray}
Note that $\beta>0$ ($<0$) implies $\rho_{\Lambda,2}<\rho_{\Lambda,1}$
($\rho_{\Lambda,1}<\rho_{\Lambda,2}$), and $C<0$ implies
$\rho_{\Lambda,2}<\rho_o<\rho_{\Lambda,1}$ for $\beta>0$
($\rho_{\Lambda,1}<\rho_o<\rho_{\Lambda,2}$ for $\beta<0$).\\

\noindent {\bf I1:} $\beta>0$, $P_{o}>0$, $\rho<\rho_{\Lambda,2}$,
hence we consider $\rho_{\Lambda,2}>0$. In this case $0 < a <
a_{1} $ with $\rho_{\Lambda,2}>\rho>-\infty$ and the fluid behaves
in a standard manner. The open models are all non-physical as they
evolve to the $\rho<0$ region of the phase space. The flat models
evolve from an expanding de Sitter phase to a contracting de
Sitter phase. The closed model region contains a generalized
Einstein static fixed point and models which
oscillate indefinitely (between minima and maxima $a$ and $\rho$).\\

\noindent {\bf I2:} $\beta>0$, $P_{o}>0$, $\rho_{\Lambda,2}<\rho
<\rho_{\Lambda,1}$. In
this case $0<a<\infty$ with $\rho_{\Lambda,2}<\rho
<\rho_{\Lambda,1}$ and the fluid behaves in a phantom manner. The
open models evolve from one expanding de Sitter phase
($\rho=\rho_{\Lambda,2}$) to more rapid (greater $\rho$ and $H$)
de Sitter phase ($\rho=\rho_{\Lambda,1}$), however the spatial
curvature is negative in the past and asymptotically approaches
zero in the future. The flat models behave in a similar manner
except that the curvature remains zero. The closed models undergo
a phantom bounce with asymptotic de Sitter behavior, that is they
evolve from a contracting de Sitter phase, reach a minimum $a$,
minimum $\rho$ and then evolve to a expanding de Sitter phase.\\

\noindent {\bf I3:} $\beta>0$, $P_{o}>0$, $\rho>\rho_{\Lambda,1}$.
In this case $a_{1} < a < \infty $ with $\infty>\rho
>\rho_{\Lambda,1}$ and the fluid behaves in a standard manner.
All flat and open models expand from a singularity at $a=a_{1}$
and asymptotically evolve to a expanding de Sitter phase
($\rho=\rho_{\Lambda,1}$). A subset of closed models evolve from a
contracting de Sitter phase to minimum $a$ (maximum $\rho$),
bounce and then evolve to an expanding de Sitter phase. Another
subset of closed models expand from a singularity at $a=a_{1}$,
reach a maximum $a$ and minimum
$\rho$, only to re-collapse.\\

\noindent {\bf I4:} $\beta<0$, $P_{o}<0$, $\rho<\rho_{\Lambda,1}$.
In this case $a_{1} < a< \infty $, with
$-\infty<\rho<\rho_{\Lambda,1}$ and the fluid behaves in a phantom
manner. The open models are all non-physical as they evolve from
the $\rho<0$ region of the phase space. The flat and closed models
evolve from a contracting de Sitter phase, bounce at minimum $a$
and $\rho$, then re-expand, asymptotically approaching a expanding
de Sitter phase.\\

\noindent {\bf I5:} $\beta<0$, $P_{o}<0$, $\rho_{\Lambda,1}<\rho
<\rho_{\Lambda,2}$ (where $\rho_{\Lambda,1}<\rho_{\Lambda,2}$). In
this case $0<a<\infty$ with $\rho_{\Lambda,2}>\rho
>\rho_{\Lambda,1}$ and the fluid behaves in a standard manner.
The open models evolve from a expanding de Sitter phase
($\rho=\rho_{\Lambda,2}$) to less rapid (lower $\rho$ and $H$) de
Sitter phase ($\rho=\rho_{\Lambda,1}$) with the spatial curvature
being negative in the past and zero asymptotically in the future.
The flat models behave in a similar manner, except that the
curvature remains zero throughout the evolution. The closed models
can undergo a phantom bounce with asymptotic de Sitter behavior in
the future and past, a  subset of these models enter a
loitering phase both before and after the bounce. There are a
subset of closed models which oscillate indefinitely.\\

\noindent {\bf I6:} $\beta<0$, $P_{o}<0$, $\rho>\rho_{\Lambda,2}$.
In this case $0 < a < a_{1} $ with $\rho_{\Lambda,2}<\rho <\infty$
and the fluid behaves in a phantom manner. All models have a
future singularity at $a=a_{1}$, with closed models contracting
from a past singularity to a minimum $a$ and $\rho$ before
re-expanding (phantom bounce).


\subsection{The Singularities}

In general, singularities may behave in qualitatively different
ways. The singularities present for the non linear EoS
are quite different from the standard ``Big Bang"/``Big Crunch"
singularity. The standard singularities are such that:
\begin{itemize}
\item  ``Big Bang"/``Big Crunch" : For $a \to 0$,
$\rho \to \infty$.
\end{itemize}

If the singularity occurs in the past (future) we refer to it as a
``Big Bang" (``Big Crunch"). In order to differentiate between
various types of singularities, we will use the following
classification system for future singularities~\cite{NOT} (cf.\ also~\cite{barrow}):
\begin{itemize}
\item  Type I (``Big Rip'') : For $t \to t_{\star}$, $a \to \infty$,
$\rho \to \infty$ and $|P| \to \infty$.
\item  Type II (``sudden'') : For $t \to t_{\star}$, $a \to a_{\star}$,
$\rho \to \rho_{\star}$ or $0$ and $|P| \to \infty$.
\item  Type III : For $t \to t_{\star}$, $a \to a_{\star}$,
$\rho \to \infty$ and $|P| \to \infty$.
\item  Type IV : For $t \to t_{\star}$, $a \to a_{\star}$,
$\rho \to \rho_{\star}$ or $0$, $|P| \to |P_{\star}|$ or $0$ and
 derivatives of $H$ diverge.
\end{itemize}

\noindent Here $t_{\star}$, $a_{\star}$, $\rho_{\star}$ and
$|P_{\star}|$ are constants with $a_{\star}\neq 0$. The main
difference in our case is that the various types of singularities
may occur in the past or the future. The future singularity
described in case {\bf A2} falls into the category of Type III,
however, the past singularity mentioned in case {\bf A1} is also a
Type III singularity. In the case of the full quadratic EoS, all
singularities which occur for a finite scale factor ($a=a_{1}$)
are of Type III.

\section{ High energy regime Dynamics}\label{sec3}

\subsection{The dimensionless  dynamical system}\label{sec3_a}
It is convenient to describe the dynamics in terms of
dimensionless variables. In the high energy regime these are:
\begin{equation}
x=\frac{\rho}{|\rho_{c}|}\;,~~ y=\frac{H}{\sqrt{|\rho_{c}|}}\;,~~
\eta=\sqrt{|\rho_{c}|}t\;.
\end{equation}

\noindent The system of equations (\ref{energycons})-(\ref{Ray}) then changes into:
\begin{eqnarray}
x'&=& -3 y \left( (\alpha +1)x + \epsilon x^2 \right),\nonumber\\
y' &=& -y^{2} - \frac{1}{6} \left( (3\alpha +1)x + 3\epsilon x^{2}
\right) ,\label{HED}
\end{eqnarray}

\noindent and the Friedman equation (\ref{Friedman}) gives
\begin{eqnarray}\label{fried_dim}
y^{2} &=& \frac{x}{3} - \frac{K}{|\rho_{c}|a^2}.
\end{eqnarray}

\noindent The discrete parameter $\epsilon$ denotes the sign of
the quadratic term, $\epsilon\in\{-1,1\}$. The primes denote
differentiation with respect to $\eta$, the normalized time
variable. The variable $x$ is the normalized energy density and
$y$ the normalized Hubble function. We will only consider the
region of the phase space for which the energy density remains
positive ($x\geq0$). The system of equations above is of the form
$u'_{i}=f_{i}(u_{j})$. Since this system is autonomous,
trajectories in phase space connect the fixed/equilibrium points
of the system ($u_{j,o}$), which satisfy the system of equations
$f_{i}(u_{j,0})=0$. The fixed points of the high energy system and
their existence conditions (the conditions for which $x\geq0$ and
$x,y\in\mathbb{R}$) are given in Table~\ref{Tab1}.

\begin{center}
\begin{table}[h!]\caption{\label{Tab1}Location and existence conditions
($x\geq0$ and $x,y\in\mathbb{R}$) of the fixed points of the high
energy regime system.}
\begin{tabular*}{0.47\textwidth}{@{\extracolsep{\fill}}cccc}
\hline \hline
Name & $x$&$y$& Existence \\
\hline
\\
$M$ & $0$ & $0$ & $-\infty<\alpha<\infty$ \\
$E$ & $-\frac{\epsilon(3\alpha+1)}{3}$ & $0$ &
$\epsilon(3\alpha+1)<0$ \\
$dS_{+}$ & $-\epsilon(\alpha+1)$ &
$+\sqrt{\frac{-\epsilon(\alpha+1)}{3}}$ &
$\epsilon(\alpha+1)<0$ \\
$dS_{-}$ & $-\epsilon(\alpha+1)$ &
$-\sqrt{\frac{-\epsilon(\alpha+1)}{3}}$&
$\epsilon(\alpha+1)<0$\\
\\
\hline \hline
\end{tabular*}
\end{table}
\end{center}

The first fixed point (M) represents an empty flat (Minkowski)
model. The parabola $y^2=x/3$ is the union of trajectories
representing  flat  models, $K=0$ in Eq.~(\ref{fried_dim}) (see
Figs.\ 1 and 3-7). The trajectories below the parabola represent
open  models ($K=-1$), while trajectories above the parabola
represent closed  models ($K=+1$). The second fixed point (E)
represents a generalized static Einstein universe. This requires
some form of inflationary matter and therefore may only exists
when $\alpha<-1/3$ if $\epsilon=+1$ and when $\alpha>-1/3$ if
$\epsilon=-1$. The last two points represent expanding and
contracting spatially flat de Sitter models ($dS_{\pm}$). These
points exist when the fluid permits an effective cosmological
constant point, $x_{\Lambda} :=\rho_\Lambda/\rho_c=
-\epsilon(\alpha+1)$; in addition $x_{\Lambda}>0$ must be true for
the fixed points to be in the physical region of the phase space.
There are further fixed points at infinity, these can be found by
studying the corresponding compactified phase space. The first
additional fixed point is at $x=y=\infty$ and represents a
singularity with infinite expansion and infinite energy density.
The second point is at $x=\infty$, $y=-\infty$ and represents a
singularity with infinite contraction and infinite energy density.

\subsection{Generalities of stability analysis}\label{sec3_b}

\indent The stability nature of the fixed points can be found by
carrying out a linear stability analysis. In brief (see
e.g.~\cite{AP} for details), this involves analyzing the behavior
of linear perturbations $u_{j}=u_{j,o}+v_{j}$ around the fixed
points, which obey the equations $v'_{i}={\bf M} v_{j}$. The
matrix ${\bf M}$ is the Jacobian matrix of the dynamical system
and is of the form:
\begin{equation}
{\bf M}_{ij}=\frac{\partial f_{i}}{\partial
u_{j}}\Bigg|_{u_{k}=u_{k,o}}.
\end{equation}

The eigenvalues $\lambda_{i}$ of the Jacobian matrix evaluated at
the fixed points tell us the linear stability character of the
fixed points. The fixed point is said to be hyperbolic if the real
part of the eigenvalues is non-zero ($\mathbb{R}(\lambda_{i})\neq0
$). If all the real parts of the eigenvalues are positive
($\mathbb{R}(\lambda_{i})>0 $) the point is said to be a repeller.
Any small deviations from this point will cause the system to move
away from this state. If all the real parts are negative
($\mathbb{R}(\lambda_{i})<0 $), the point is said to be an
attractor. This is because if the system is perturbed away from
this state, it will rapidly return to the equilibrium state. If
some of the values are positive, while others are negative then
the point is said to be a saddle point. If the eigenvalues of the
fixed point are purely imaginary then the point is a center. If
the center nature of the fixed point is confirmed by some
non-linear analysis, then the trajectories will form a set of
concentric closed loops around the point. If the eigenvalues do
not fall into these categories, we will resort to numerical
methods to determine their stability.

The eigenvalues for the fixed points of the system
(Eq.'s~(\ref{HED})) are given in Table~\ref{Tab2} and the linear
stability character is given in Table~\ref{Tab3}.

\begin{center}
\begin{table}[h!]\caption{\label{Tab2}Eigenvalues for
the fixed points of the high energy regime system.}
\begin{tabular*}{0.47\textwidth}{@{\extracolsep{\fill}}ccc}
\hline \hline
Name & $\lambda_{1}$ & $\lambda_{2}$ \\
\hline
\\
$M$ & $0$ & $0$ \\
$E$ & $\sqrt{\epsilon}\frac{(3\alpha+1)}{3}$ & $-\sqrt{\epsilon}
\frac{(3\alpha+1)}{3}$ \\
$dS_{+}$ & $(\alpha+1)\sqrt{-3\epsilon(\alpha+1)}$ &
$-\frac{2}{3}\sqrt{-3\epsilon(\alpha+1)}$ \\
$dS_{-}$ & $-(\alpha+1)\sqrt{-3\epsilon(\alpha+1)}$ &
$\frac{2}{3}\sqrt{-3\epsilon(\alpha+1)}$ \\
\\
\hline \hline
\end{tabular*}
\end{table}
\end{center}

\begin{center}
\begin{table}[h!]\caption{\label{Tab3} The linear
stability of the fixed points for the high energy regime system.}
\begin{tabular*}{0.47\textwidth}{@{\extracolsep{\fill}}ccc}
\hline \hline
Name & $\epsilon=+1$ & $\epsilon=-1$ \\
\hline
\\
$M$ & undefined & undefined \\
$E$ & Saddle ($\alpha \neq -1/3$) &
Center ($\alpha \neq -1/3$) \\
$dS_{+}$ & Attractor ($\alpha < -1$)  &
Saddle ($\alpha > -1$) \\
$dS_{-}$ & Repeller ($\alpha < -1$)  &
Saddle ($\alpha > -1$)  \\
\\
\hline \hline
\end{tabular*}
\end{table}
\end{center}

\subsection{The $\epsilon=+1$ case}\label{sec3a}

We first consider the system when we have a positive quadratic
energy density term ($\epsilon=+1$) in the high energy regime EoS.
We will concentrate on the region around the origin as this is
where the finite energy density fixed points are all located. The
plots have been created using the symbolic mathematics application
Maple 9.5. The individual plots are made up by three layers, the
first is a directional (represented by grey arrows) field plot of
the state space. The second layer represents the separatrices and
fixed points of the state space . A separatrix (black lines) is a
union of trajectories that marks a boundary between subsets of
trajectories with different properties and can not be crossed. The
fixed points are represented by black dots. The final layer
represents some example trajectories (grey lines) which have been
calculated by numerically integrating the system of equations for
a set of initial conditions. The character of the fixed point M is
undefined and so is determined numerically. The fixed point
representing the generalized Einstein static solution is a saddle
point. The fixed points representing the generalized expanding
(contracting) de Sitter points are attractor (repeller) points.
The trajectories or fixed points in the $y>0$ ($y<0$) region
represent expanding (contracting) models. We will mainly discuss
the right hand side of the state space (expanding models) as in
general the corresponding trajectory on the left hand side is
identical under time reversal.

\subsubsection{The $\alpha<-1$ sub-case}\label{sec3a1}

\begin{figure}[t!]
\begin{center}
\hspace{0.4cm}\includegraphics[width=8.5cm,height=8.5cm,angle=270]{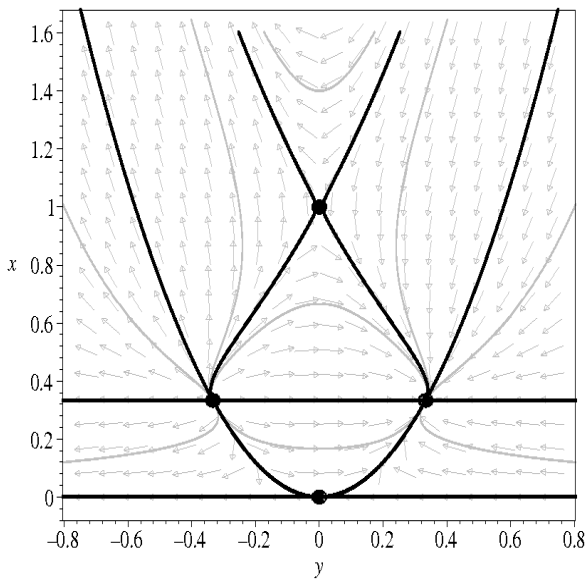}
\caption{The phase space for the high energy regime system with
$\epsilon=+1$ and $\alpha<-1$. The upper (lower) region
corresponds to the case {\bf B1} ({\bf C1}).}
\end{center}
\label{fig1}
\end{figure}

The phase space of the system is considered when $\alpha<-1$ and
is shown in Fig. 1. The lowest horizontal line ($x=0$) is the
separatrix for open models ($K=-1$) and will be referred to as the
open Friedmann separatrix (OFS). The trajectories on the
separatrix represent Milne models ($x=0$, $K=-1$ and
$a(\eta)\propto\eta$) which are equivalent to a Minkowski
space-time in a hyperbolic co-ordinate system. The second higher
horizontal line ($x_{\Lambda}=-(\alpha+1)$) is the separatrix
which is the dividing line between regions of phantom
($x<x_{\Lambda}$) and non-phantom/standard
behavior($x>x_{\Lambda}$), we will call this the phantom
separatrix (PS). The standard region corresponds to the case {\bf
B1}, while the phantom region corresponds to the case {\bf C1}. In
the phantom region the fluid violates the Null Energy Condition
($\rho+P<0$). This means the energy density is increasing
(decreasing) in the future for an expanding (contracting)
universe. In the standard case of the linear EoS in GR, this
occurs when ${\it w}<-1 $ and ultimately leads to a Type I
singularity~\cite{BLJM, CKW}. The parabola ($y^2=x/3$) represents
the separatrix for flat Friedmann models ($K=0$), we will call
this the flat Friedmann separatrix (FFS). The inner most thick
curve is the separatrix for closed Friedmann models ($K=+1$) and
will be called the closed Friedmann separatrix (CFS). The
separatrix has the form:
\begin{equation}
y^2 = \frac{x}{3} - \left[ \frac{A(\alpha+1)x}{(\alpha+1)+\epsilon
x} \right]^{\frac{2}{3(\alpha+1)}}.
\end{equation}

\noindent The constant $A$ is fixed by ensuring that the saddle
fixed point coincides with the fixed point representing the
generalized Einstein static model ($E$). The constant is given in
terms of the EoS parameters and has the form:
\begin{equation}
A=-\frac{2}{\epsilon(3\alpha+1)(\alpha+1)}\left(
-\frac{\epsilon(3\alpha+1)}{9}\right)^{\frac{3(\alpha+1)}{2}}.
\end{equation}

The Minkowski fixed point is located at the intersection of the
OFS and FFS. The generalized flat de Sitter fixed points are
located at the intersection of the PS and FFS. The generalized
Einstein static fixed point is located on the CFS. The
trajectories between the OFS and the PS ($0 < x < x_{\Lambda}$)
represent models which exhibit phantom behavior (the case {\bf
C1}). The open models in the phantom region are asymptotic to a
Milne model in the past and to a generalized flat de Sitter model
($dS_{+}$) in the future. The closed models in the phantom region
evolve from a contracting de Sitter phase, through a phantom phase
to an expanding de Sitter phase (phantom bounce). It is
interesting to note that unlike the standard GR case the phantom
behavior does not result in a Type I singularity but
asymptotically evolves to a expanding de Sitter phase. This is
similar to the behavior seen in the phantom generalized Chaplygin
gas case~\cite{BLJM}. The trajectories on the PS all represent
generalized de Sitter models ($x'=0$). The fixed points represent
generalized flat de Sitter models ($K=0$). The open model on the
PS represent generalized open de Sitter models ($K=-1$) in
hyperbolic co-ordinates. The closed models on the PS evolve from a
contracting phase to an expanding phase and represent generalized
closed de Sitter models ($K=+1$). The Friedmann equation can be
solved for such models to give:
\begin{equation}
\begin{array}{ll}
a(\eta) = \sqrt{\frac{3}{x_{o}}} \cosh\left[
\sqrt{\frac{x_{o}}{3}}(\eta-\eta_o)\right]~~ & \mbox{for $k=1$}\,, \\
\\
a(\eta) = \mbox{e}^{\sqrt{\frac{x_{o}}{3}}(\eta-\eta_o)}
& \mbox{for $k=0$}\,, \\
\\
a(\eta) = \sqrt{\frac{3}{x_{o}}} \sinh\left[
\sqrt{\frac{x_{o}}{3}}(\eta-\eta_o)\right] & \mbox{for $k=-1$}\,,
\end{array}
\end{equation}

The region above the PS represents models which evolve in a
non-phantom/standard manner (the case {\bf B1}). The trajectories
in the expanding region ($y>0$) of the phase space are asymptotic
to a Type III singularity in the past\footnote{The Type III
singularity appears to be a generic feature of the high energy
regime EoS and can occur both in the future and the past.}. The
trajectories outside the FFS represent open models which evolve
from a Type III singularity to a flat de Sitter phase, as do the
trajectories on the FFS. The trajectories in between the CFS and
the FFS evolve from a Type III singularity to a flat expanding de
Sitter phase but may enter a phase of loitering. Loitering is
characterized by the Hubble parameter dipping to a low value over
a narrow red-shift range, followed by a rise again. In order to
see this more clearly, we have plotted the normalized Hubble
parameter ($y$) as a function of scale factor for three different
trajectories in Fig. 2.
\begin{figure}[t!]
\begin{center}
\hspace{0.4cm}\includegraphics[width=8.5cm,height=8.5cm,angle=270]{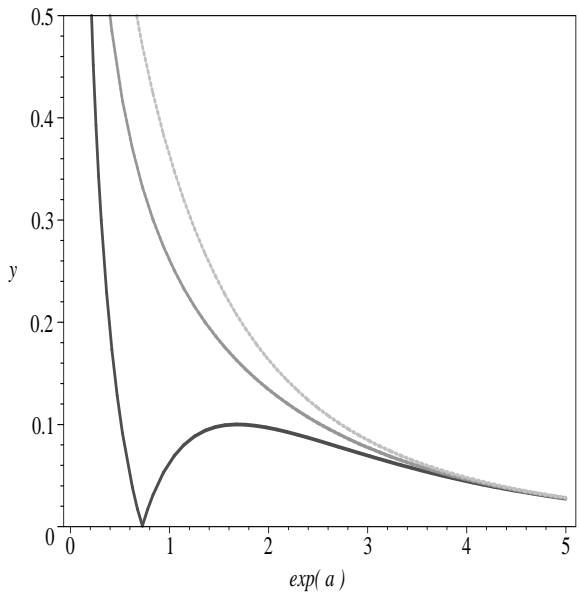}
\caption{The normalized Hubble parameter, $y$ for models with
differing curvature. Starting from the top we have open, flat and
the closed models. }
\end{center}
\label{fig2}
\end{figure}
The top two curves represent the open and flat models, with the
Hubble parameter dropping off quicker for the flat Friedmann
model. The lower most curve is the Hubble parameter for the closed
model. The plot shows that the closed model evolves to a loitering
phase. Loitering cosmological models in standard cosmology were
first found for closed FLRW models with a cosmological constant.
More recently, brane-world models which loiter have been found
~\cite{SS}, these models are spatially flat but can behave
dynamically like a standard FLRW closed model. The interesting
point here is that the models mentioned above loiter without the
need of a cosmological constant (due to the appearance of an
effective cosmological constant), the topology is asymptotically
closed in the past and flat in the future. The trajectories inside
the CFS can have two distinct types of behavior corresponding to
the central regions above and below the generalized Einstein
static fixed point. Trajectories in the lower region represent
closed models which evolve from a contracting de Sitter phase,
bounce and then evolve to a expanding de Sitter phase. The
trajectories in the upper region evolve from a Type III
Singularity, expand to a maximum $a$ (minimum $x$) and then
re-collapse to a Type III singularity (we will refer to such
re-collapsing models as turn-around models).

\subsubsection{The $-1<\alpha<-1/3$ sub-case}\label{sec3a2}

\begin{figure}[t!]
\begin{center}
\hspace{0.4cm}\includegraphics[width=8.5cm,height=8.5cm,angle=270]{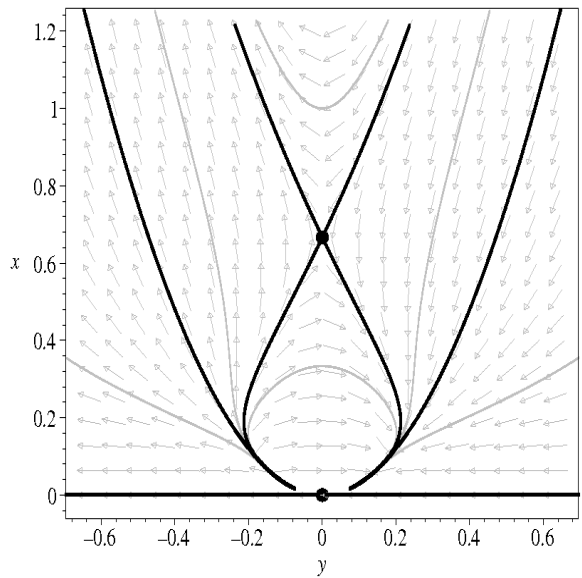}
\caption{The phase space for the high energy regime system with
$\epsilon=+1 $ and $-1<\alpha<-1/3$. The entire region corresponds
to the case {\bf A1}.}
\end{center}
\label{fig3}
\end{figure}

The phase space for the system when $-1<\alpha<-1/3$ is shown in
Fig. 3. The fixed points representing the flat generalized de
Sitter models are no longer in the physical region ($x>0$) of the
phase space. The open, flat and closed Friedmann separatrices
(OFS, FFS and CFS) remains the same. The phantom separatrix (PS)
is no longer present and all trajectories represent models with
non-phantom/standard fluids (this corresponds to the case {\bf
A1}). The main difference is that the generic future attractor is
now the Minkowski model. The trajectories between the OFS and FFS
now evolve from a Type III singularity to a Minkowski model, as do
the flat Friedmann models. The models between the FFS and CFS now
evolve from a Type III singularity to a Minkowski with the
possibility of entering a loitering phase (as before the model is
asymptotically flat in the future). The trajectories inside the
CFS and above the Einstein static fixed point still represent
turn-around models. The trajectories inside the OFS and below the
Einstein static model now represent standard bounce models, that
is they evolve from a Minkowski model, contract to a finite size,
bounce and then expand to a Minkowski model.

\subsubsection{The $\alpha\geq-1/3$ sub-case}\label{sec3a3}

\begin{figure}[t!]
\begin{center}
\hspace{0.4cm}\includegraphics[width=8.5cm,height=8.5cm,angle=270]{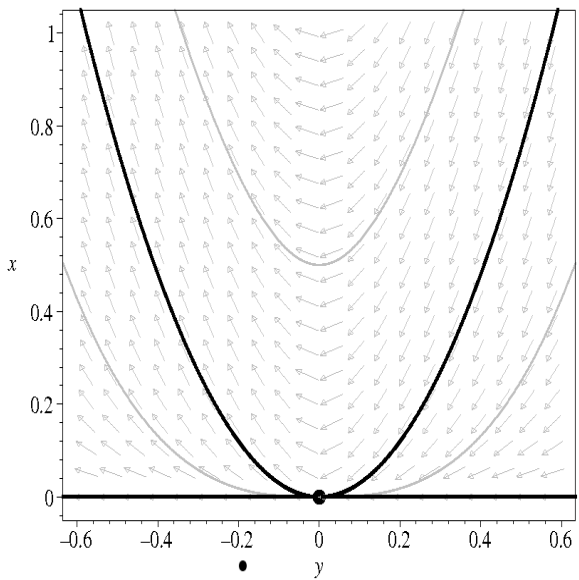}
\caption{The phase space for the high energy regime system with
$\epsilon=+1$ and $\alpha\geq-1/3$. The entire region corresponds
to the case {\bf A1}.}
\end{center}
\label{fig4}
\end{figure}

Next we consider the system when $\alpha\geq-1/3$, the phase space
is shown in Fig. 4. The fixed point representing the Einstein
static models is now located in the $x<0$ region of the phase
space. The fluid in the entire physical region behaves in a
non-phantom manner and corresponds to the case {\bf A1}. The OFS
and FFS remain the same and the CFS is no longer present. The
trajectories between the OFS and FFS evolve from a Type III
singularity to a Minkowski model. All trajectories above the FFS
now represent turn-around models which start and terminate at a
Type III singularity. The behavior of the models is qualitatively
the same as that of the standard FLRW model with a linear EoS
where ${\it w}=\alpha$, in the linear EoS case the Type III
singularity is replaced by a standard ``Big Bang".

\subsection{The $\epsilon=-1$ case}\label{sec3b}

We now consider the system when we have a negative quadratic
energy density term ($\epsilon=-1$) in the high energy regime EoS.
The character of the fixed point M is still undefined. The fixed
point representing the generalized Einstein static model is now a
center. The fixed points representing the expanding/contracting
flat de Sitter points now have saddle stability. As before, the
trajectories or fixed points in the $y>0$ ($y<0$) region represent
expanding models (contracting models), the black lines represent
separatrix, grey lines represent example trajectories and fixed
points are represented by black dots.

\subsubsection{The $\alpha<-1$ sub-case}\label{sec3b1}

\begin{figure}[t!]
\begin{center}
\hspace{0.4cm}\includegraphics[width=8.5cm,height=8.5cm,angle=270]{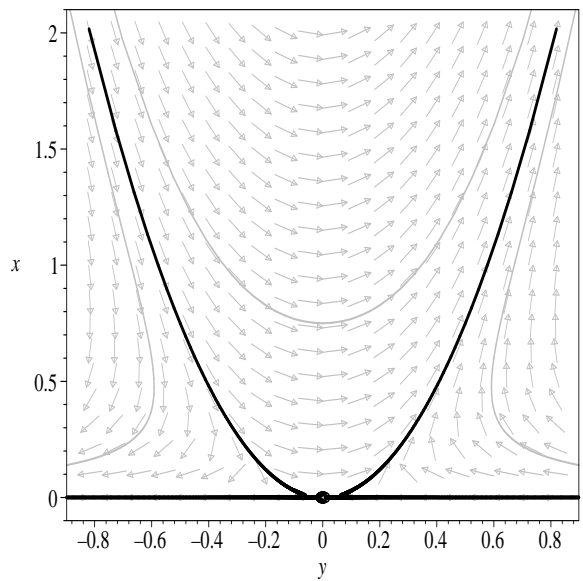}
\caption{The phase space for the high energy regime system with
$\epsilon=-1$ and $\alpha<-1$. The entire region corresponds to
the case {\bf A2}. }
\end{center}
\label{fig5}
\end{figure}

The phase space of the system when $\alpha<-1$ is shown in Fig. 5.
The horizontal line ($x=0$) is still the open Friedman separatrix
(OFS). The parabola is the flat Friedmann separatrix (FFS). The
intersection of the OFS and FFS coincides with the Minkowski fixed
point. All the trajectories in the physical region of the phase
space exhibit phantom behavior (corresponding to the case {\bf
A2}), the energy density increases in an expanding model. The
trajectories in the expanding (contracting) region in general
evolve to a Type III singularity in the future (past). The
trajectories between the OFS and the FFS are asymptotic to a Milne
model in the past and are asymptotic to a Type III singularity in
the future. The trajectories on the FFS start from a Minkowski
model and enter a phase of super-inflationary expansion and evolve
to a Type III singularity. Trajectories that start in a
contracting phase during which the energy density decreases, reach
a minimum $a$ (minimum $x$) and then expand where the energy
density increases represent phantom bounce models. The
trajectories above the FFS represent closed models which evolve
through a phantom bounce, but start and terminate in a Type III
singularity. The behavior of the models is qualitatively the same
as that of the FLRW models with a phantom linear EoS (${\it
w}<-1$) except that the Type III singularity is replaced by a Type
I (``Big Rip") singularity.

\subsubsection{The $-1<\alpha<-1/3$ sub-case}\label{sec3b2}

\begin{figure}[h!]
\begin{center}
\hspace{0.4cm}\includegraphics[width=8.5cm,height=8.5cm,angle=270]{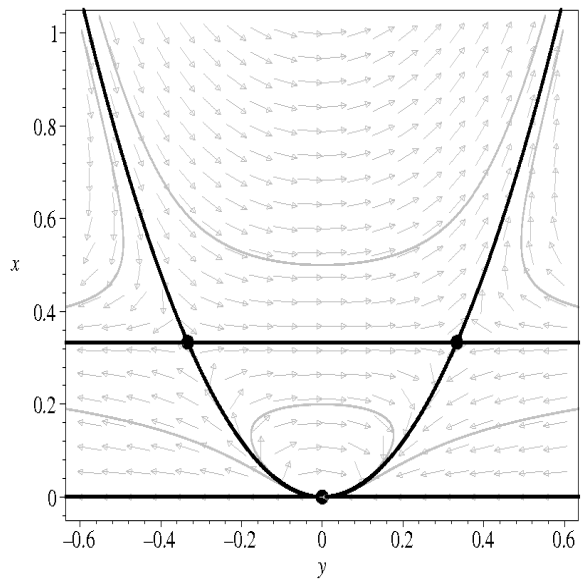}
\caption{The phase space for the high energy regime system with
$\epsilon=-1 $ and $-1<\alpha<-1/3$. The upper (lower) region
corresponds to the case {\bf B2} ({\bf C2}). }
\end{center}
\label{fig6}
\end{figure}

The phase space for the system when $-1<\alpha < -1/3$ is shown in
Fig. 6. The lowest horizontal line ($x=0$) is the OFS. The second
higher horizontal line, $x_{\Lambda} = (\alpha+1) $ is the phantom
separatrix (PS), this divides the state space into regions of
phantom ($x>x_{\Lambda}$) and non-phantom/standard behavior
($x<x_{\Lambda}$). The phantom region corresponds to the case {\bf
B2} and the standard region corresponds to the case {\bf C2}. The
flat de Sitter ($dS_{\pm}$) points are located at the intersection
of the FFS and the PS. The open models in the standard matter
region ($0<x<x_{\Lambda}$) are past asymptotic to open expanding
de Sitter models in the past and evolve to Minkowski models in the
future. The closed models in the region represent the standard
bounce models, that is they evolve from a Minkowski model,
contract to a minimum $a$ (maximum $x$) and then expand to a
Minkowski model. The trajectories above the PS ($x>(\alpha+1)$)
all exhibit phantom behavior. The open models in this region are
past asymptotic to open de Sitter models in the past and evolve to
a Type III singularity in the future. The closed models in the
region all represent models which undergo a phantom bounce but
start and terminate in a Type III singularity.

\subsubsection{The $\alpha\geq-1/3$ sub-case}\label{sec3b3}

\begin{figure}[h!]
\begin{center}
\hspace{0.4cm}\includegraphics[width=8.5cm,height=8.5cm,angle=270]{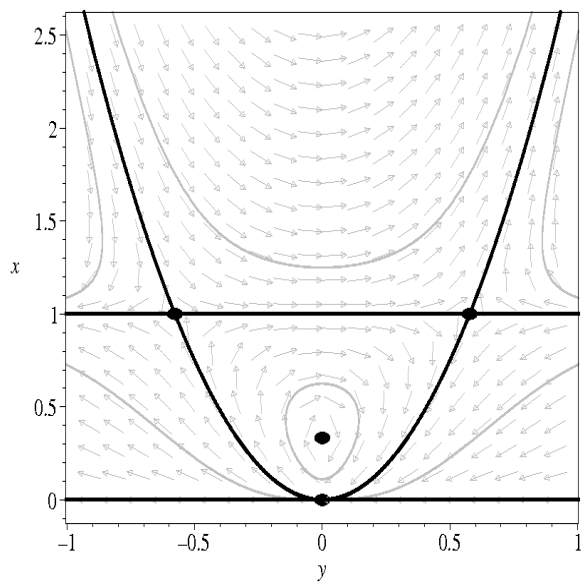}
\caption{The phase space for the high energy regime system with
$\epsilon=-1$ and $\alpha\geq-1/3$. The upper (lower) region
corresponds to the case {\bf B2} ({\bf C2}).}
\end{center}
\label{fig7}
\end{figure}

We now consider the system when $\alpha\geq-1/3$, the phase space
is shown in Fig. 7. The OFS, FFS and the PS are all still present,
the phantom regions still corresponds to the case {\bf B2} and the
standard region to the case {\bf C2}. The trajectories in the
phantom region ($x>x_{\Lambda}$) behave in a similar manner to the
previous case, as do the open models in the standard matter region
($0<x<x_{\Lambda}$). The main difference is in the region
representing closed models ($K=1$) with non-phantom behavior.
There is now a new fixed point which represents a generalized
Einstein static model ($E$). The closed models in the region now
represent oscillating models. This is represented by closed
concentric loops centered on the Einstein static fixed point.
These oscillating models also appear in the low energy system and
will be discussed in more detail later.

\section{Low energy regime Dynamics }\label{sec4}

\subsection{The dimensionless dynamical system}
\noindent We now consider the system of equations for the low
energy regime EoS, which can be simplified and expressed in terms
of the following dimensionless variables:
\begin{equation}
x=\frac{\rho}{|P_{o}|}\;,~~ y=\frac{H}{\sqrt{|P_{o}|}}\;,~~
\eta=\sqrt{|P_{o}|}t\;.
\end{equation}

\noindent The system of equations is then:
\begin{eqnarray}
y^{2} &=& \frac{x}{3} - \frac{K}{|P_{o}|a^2}, \\
y' &=& -y^{2} - \frac{1}{6} \left( 3\epsilon_p    + (3\alpha +1)x  \right) ,\\
x'&=& -3 y \left( \epsilon_p + (\alpha +1)x  \right).
\end{eqnarray}

\noindent The discrete parameter $\epsilon_p$ denotes the sign of
the pressure term, $\epsilon_p\in\{-1,1\}$. The primes denote
differentiation with respect to the new $\eta$. The variables $x$
and $y$ are the new normalized energy density and Hubble
parameter. As before only the positive energy density region of
the phase space will be considered. The fixed points of the system
and the existence conditions are given in Table~\ref{Tab4}. As
before, by existence we mean the conditions on the parameters to
insure $x\geq0$ and $x,y\in\mathbb{R}$.
\begin{center}
\begin{table}[h!]\caption{\label{Tab4}Location and existence conditions of the
fixed points of the low energy regime system.}
\begin{tabular*}{0.47\textwidth}{@{\extracolsep{\fill}}cccc}
\hline \hline
Name & $x$ & $y$ &  Existence  \\
\hline
\\
$E$ & $-\frac{3\epsilon_p }{(3\alpha +1)}$ & $0$ &
$\frac{\epsilon_p}{(3\alpha+1)}<0$ \\
$dS_{+}$ & $-\frac{\epsilon_p}{(\alpha+1)}$ &
$+\sqrt{\frac{-\epsilon_p}{3(\alpha+1)}}$ &
$\frac{\epsilon_p}{(\alpha+1)}<0$ \\
$dS_{-}$ & $-\frac{\epsilon_p}{(\alpha+1)}$ &
$-\sqrt{\frac{-\epsilon_p}{3(\alpha+1)}}$ &
$\frac{\epsilon_p}{(\alpha+1)}<0$ \\
\\
\hline \hline
\end{tabular*}
\end{table}
\end{center}

The Minkowski model ($x=y=0$) is no longer a fixed point of the
system. The first fixed point (E) represents a generalized static
Einstein model. This requires the overall effective equation of
state to be that of inflationary matter and therefore only exists
when $\epsilon_p/(3\alpha+1)<0$. The last two points represent
generalized expanding and contracting flat de Sitter models. These
points only exist if the fluid permits an effective cosmological
constant point $\tilde{x}_{\Lambda}=\tilde{\rho}_{\Lambda}/
{|P_{o}|} = -\epsilon_p /(\alpha+1)$, also
$\tilde{x}_{\Lambda}\geq0$ for the points to be in the physical
region of the phase space. The eigenvalues of the equilibrium
points are given in Table~\ref{Tab5}, while the linear stability
character is given in Table~\ref{Tab6}.
\begin{center}
\begin{table}[h!]\caption{\label{Tab5}Eigenvalues of the
fixed points of the low energy regime system.}
\begin{tabular*}{0.47\textwidth}{@{\extracolsep{\fill}}ccc}
\hline \hline
Name & $\lambda_{1}$ & $\lambda_{2}$  \\
\hline
\\
$E$ & $\sqrt{-\epsilon_p}$ & $-\sqrt{-\epsilon_p}$ \\
$dS_{+}$ & $ \sqrt{\frac{-3(\alpha+1)}{\epsilon_p}} $ & $
-\frac{2}{\sqrt{-3\epsilon_p(\alpha+1)}} $ \\
$dS_{-}$ & $ -\sqrt{\frac{-3(\alpha+1)}{\epsilon_p}} $ & $
\frac{2}{\sqrt{-3\epsilon_p(\alpha+1)}}$ \\
\\
\hline \hline
\end{tabular*}
\end{table}
\end{center}

\begin{center}
\begin{table}[h!]\caption{\label{Tab6} The linear
stability of the fixed points for the low energy regime system.}
\begin{tabular*}{0.47\textwidth}{@{\extracolsep{\fill}}ccc}
\hline \hline
Name & $\epsilon_p=+1$ & $\epsilon_p=-1$  \\
\hline
\\
$E$ & Center ($\alpha \neq -1/3$) & Saddle ($\alpha \neq -1/3$) \\
$dS_{+}$ & Saddle ($\alpha < -1$)  &
Attractor ($\alpha > -1$)  \\
$dS_{-}$ & Saddle ($\alpha < -1$)  &
Repeller ($\alpha > -1$)  \\
\\
\hline \hline
\end{tabular*}
\end{table}
\end{center}

\subsection{The $\epsilon_p=+1$ case}\label{sec4a}

We start by considering the system when we have a positive
constant pressure term ($\epsilon_p=+1$) in the low energy regime
EoS. The Minkowski ($x=y=0$) point is no longer present and the
Einstein static solution has the stability character of a center.
The fixed points representing the generalized
expanding/contracting de Sitter points ($dS_\pm$) now have saddle
stability. As before black lines represent separatrix, grey lines
represent example trajectories and black dots represent fixed
points of the system.

\subsubsection{The $\alpha<-1$ sub-case}\label{sec4a1}

\begin{figure}[h!]
\begin{center}
\hspace{0.4cm}\includegraphics[width=8.5cm,height=8.5cm,angle=270]{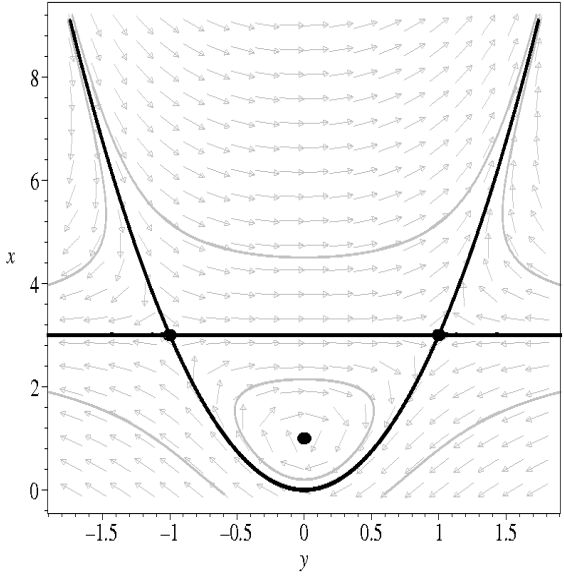}
\caption{The phase space for the low energy regime system with
$\epsilon_p=+1$ and $ \alpha < -1$. The upper (lower) region
corresponds to the case {\bf E1} ({\bf F1}).}
\end{center}
\label{fig8}
\end{figure}

The phase space for the system when $ \alpha < -1$ is shown in
Fig. 8. The open Friedmann separatrix ($x=0$) is no longer
present, and the $x=y=0$ point is no longer a fixed point of the
system. The horizontal line ($\tilde{x}_\Lambda=-(\alpha+1)^{-1}$)
is the phantom separatrix (PS), dividing the state space into
regions with phantom ($x>\tilde{x}_\Lambda$) and standard
($x<\tilde{x}_\Lambda$) behavior. The phantom region corresponds
to the case {\bf E1} and the standard region corresponds to the
case {\bf F1}. The parabola $y^2 = x/3 $ is the separatrix
representing the flat Friedmann models (FFS), this divides the
remaining trajectories into open and closed models. The
intersection of the PS and FFS coincides with the fixed points of
the generalized flat de Sitter models.

The trajectories in the upper region that start in a contracting
phase (during which the energy density decreases), reach a minimum
$a$ (minimum $x$) and then expand, representing phantom bounce
models which terminate in a Type I singularity. The closed models
in the phantom region ($x>\tilde{x}_\Lambda$) represent phantom
bounce models which start and terminate in a Type I singularity
\footnote{The Type I singularity is a generic feature of the low
energy regime EoS and can appear both in the future and the
past.}. The open models in the phantom region are asymptotic to
open de Sitter models in the past and evolve to a Type I
singularity in the future. The trajectories below the PS
($x<\tilde{x}_\Lambda$) represent models which all behave in a
standard manner (the {\bf F1} case). The open models in this
region are all non-physical as they all evolve to the $x<0$ region
of the phase space. The region corresponding to closed models
(above the FFS) contain a fixed point which represents the
generalized Einstein static (E) model. The region is filled by a
infinite set of concentric closed loops centered on the Einstein
static fixed point, the closed loops represent oscillating models.
The Friedmann equation for such models is given by:
\begin{equation}
y^2 = \frac{x}{3} - K \left[ \frac{\epsilon_p
+(\alpha+1)x}{B(\alpha+1)} \right]^{\frac{2}{3(\alpha+1)}}.
\end{equation}

\noindent The constant $B$ is fixed by the location of the
Einstein fixed point ($E$). The constant is given in terms of
$\alpha$ and $\epsilon_p$:
\begin{equation}
B=\frac{-2\epsilon_p}{(\alpha+1)(3\alpha+1)}\left(
\frac{3\alpha+1}{-\epsilon_p}\right)^{\frac{3(\alpha+1)}{2}}.
\end{equation}

These oscillating models appear for $-\infty<\alpha<-1/3$ when
$\epsilon_p=+1$ and are qualitatively similar to the oscillating
models seen in the high energy case. The exact behavior of the
variables for these models can be calculated by fixing the EoS
parameter $\alpha$. The qualitative behavior remains the same for
the models, however the maximum and minimum values of the
variables change. In the case of $\alpha=-2/3$ the equations are
greatly simplified, the scale factor oscillates such that:
\begin{equation}
a(\eta)= a_{o}\left( 1 + \sqrt{{1-K}} \sin(\eta_{o}-\eta) \right)
\end{equation}

\noindent The maximum and minimum scale factor is then:
\begin{equation}
a_{max}=a_{o}( 1 + \sqrt{{1-K}})\;,~~ a_{min}=a_{o}( 1 -
\sqrt{{1-K}})\;.
\end{equation}

\noindent The normalized hubble parameter ($y$) is:
\begin{equation}
y=y_{o}\frac{ \sqrt{{1-K}} \cos(\eta_{o}-\eta)}{ 1 + \sqrt{{1-K}}
\sin(\eta_{o}-\eta) }.
\end{equation}

\noindent The maximum and minimum $y$ is given by:
\begin{equation}
y_{max}=y_{o}\sqrt{\frac{1-K}{K}}\;,~~
y_{min}=-y_{o}\sqrt{\frac{1-K}{K}}\; .
\end{equation}

\noindent The normalized energy density ($x$) is given by:
\begin{equation}
x=x_{o} \left( \frac{ 1 -\sqrt{{1-K}} \sin(\eta_{o}-\eta)}{ 1 +
\sqrt{{1-K}} \sin(\eta_{o}-\eta) } \right).
\end{equation}

\noindent The maximum and minimum $x$ are:
\begin{equation}
x_{max}=x_{o}\left( \frac{ 1 +\sqrt{{1-K}}}{ 1 - \sqrt{{1-K}} }
\right)\;,~~ x_{min}=x_{o}\left( \frac{ 1 -\sqrt{{1-K}}}{ 1 +
\sqrt{{1-K}} } \right)\;.
\end{equation}

\subsubsection{The $-1 < \alpha< -1/3$ sub-case}\label{sec4a2}

\begin{figure}[h!]
\begin{center}
\hspace{0.4cm}\includegraphics[width=8.5cm,height=8.5cm,angle=270]{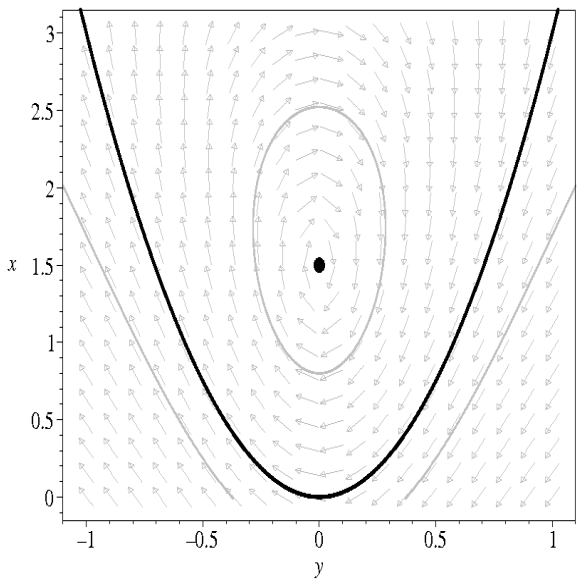}
\caption{The phase space for the low energy regime system with
$\epsilon_p=+1$ and $-1 \leq \alpha < -1/3$. The entire region
corresponds to the case {\bf D1}.}
\end{center}
\label{fig9}
\end{figure}

We now consider the case when $-1<\alpha< -1/3$, the phase space
is shown in Fig. 9. All trajectories in the physical region of the
phase space exhibit standard behavior and correspond to the case
{\bf D1}. There is only one fixed point in the $x\geq0$ region of
the phase space, this point represents the generalized Einstein
static model ($E$). The FFS represent models which evolve from a
standard ``Big Bang", evolve to a Minkowski model and then to a
standard ``Big Crunch" (turn around model). The open models (below
the FFS) are non-physical as the evolve into the $x<0$ region. The
trajectories above the separatrix represent closed models ($K>0$)
which oscillate indefinitely between a maximum and minimum $a$
(minimum and maximum $x$), as seen in the previous case.

\subsubsection{The $-1/3 \leq \alpha $ sub-case}\label{sec4a3}

\begin{figure}[h!]
\begin{center}
\hspace{0.4cm}\includegraphics[width=8.5cm,height=8.5cm,angle=270]{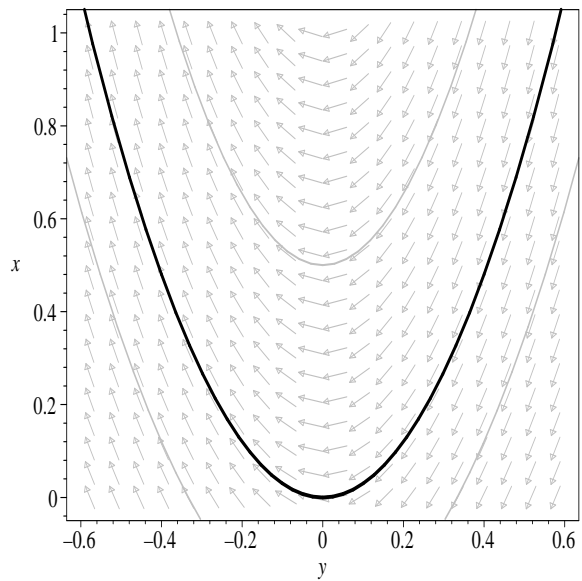}
\caption{The phase space for the low energy regime system with
$\epsilon_p=+1$ and $-1/3 \leq \alpha $. The entire region
corresponds to the case {\bf D1}. }
\end{center}
\label{fig10}
\end{figure}

The phase space for the system when $-1/3 \leq \alpha $ is shown
in Fig. 10. As in the previous sub-case the fluid behaves in a
standard manner and corresponds to the case {\bf D1}. There are no
fixed points in the physical region of the phase space. The
parabola is the FFS and represents flat model which evolve from a
``Big Bang", approach a Minkowski model and then re-collapse
(turn-around models) to a ``Big Crunch". The open models (below
the FFS) are all non-physical as they evolve to the negative
energy density region ($x<0$) of the phase space. The closed
models evolve from a ``Big Bang", reach a maximum $a$ (minimum
$x$) and re-collapse to a ``Big Crunch".

\subsection{The $\epsilon_p=-1$ case}\label{sec4b}

We now consider the system when we have a negative constant
pressure term ($\epsilon_p=-1$) in the low energy regime EoS. As
before,the Minkowski ($x=y=0$) point is no longer a fixed point of
the system and the OFS is not present. The fixed point
representing the generalized Einstein static model ($E$) has the
stability character of a saddle. The fixed points representing the
generalized expanding (contracting) flat de Sitter points now have
attractor (repeller) stability.

\subsubsection{The $\alpha < -1$ sub-case}\label{sec4b1}

\begin{figure}[h!]
\begin{center}
\hspace{0.4cm}\includegraphics[width=8.5cm,height=8.5cm,angle=270]{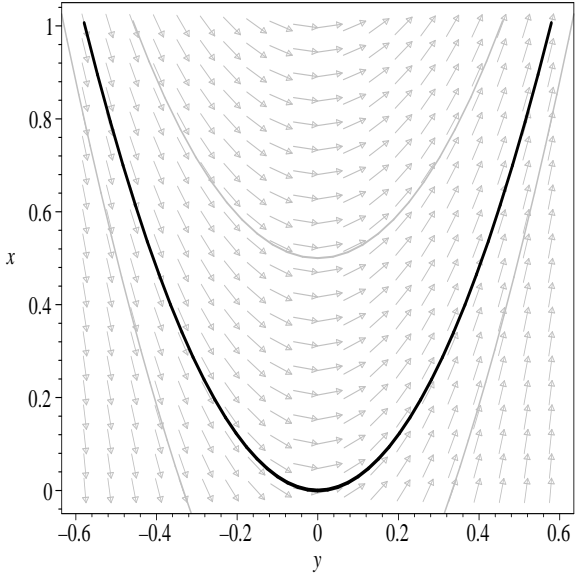}
\caption{The phase space for the low energy regime system with
$\epsilon_p=-1$ and $\alpha < -1$. The entire region corresponds
to the case {\bf D2}.}
\end{center}
\label{fig11}
\end{figure}

The phase space for the low energy system when $\alpha < -1$ is
shown in Fig. 11. All the trajectories in the $x>0$ region of the
phase space now exhibit phantom behavior and correspond to the
case {\bf D2}. The open models are all non-physical as they all
evolve from the negative energy density region of the phase space.
The flat and closed models represent phantom bounce models which
start and end in a Type I singularity. They evolve from a Type I
singularity, contract, bounce at a minimum $a$ (minimum $x$) and
expand to a Type I singularity.

\subsubsection{The $-1< \alpha < -1/3$ sub-case}\label{sec4b2}

\begin{figure}[h!]
\begin{center}
\hspace{0.4cm}\includegraphics[width=8.5cm,height=8.5cm,angle=270]{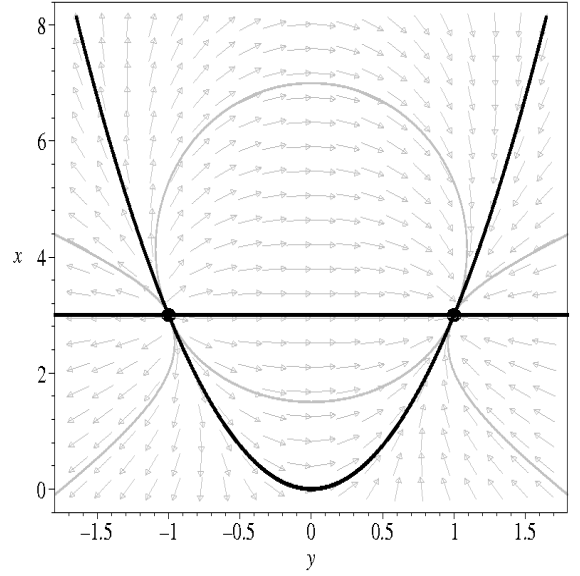}
\caption{The phase space for the low energy regime system with
$\epsilon_p=-1$ and $-1 \leq \alpha < -1/3$. The upper (lower)
region corresponds to the case {\bf E2} ({\bf F2}).}
\end{center}
\label{fig12}
\end{figure}

The phase space for the system when $-1<\alpha < -1/3$ is shown in
Fig. 12. The horizontal line, $\tilde{x}_\Lambda =
(\alpha+1)^{-1}$ is the phantom separatrix (PS), dividing the
state space into regions with phantom ($x<\tilde{x}_\Lambda$) and
standard behavior ($x>\tilde{x}_\Lambda$). The standard region
corresponds to the case {\bf E2} and the phantom region
corresponds to the case {\bf F2}. The intersection of the PS and
FFS coincides with the fixed points of the generalized flat de
Sitter models ($dS_{\pm}$). The flat expanding (contracting) de
Sitter model is the generic future attractor (repeller). The open
models in the standard matter region ($x>\tilde{x}_\Lambda$)
evolve from a standard ``Big Bang" to a flat expanding de Sitter
phase. The closed models in this region evolve from a contracting
flat de Sitter phase, reach a minimum $a$ (maximum $x$), bounce
and then evolve to expanding flat de Sitter phase. These models
represent standard bounce models with asymptotic de Sitter
behavior. The open models in the phantom region ($
x<\tilde{x}_\Lambda$) are all non-physical. The flat and closed
models in this region represent models exhibiting phantom bounce
behavior which avoid the ``Big Rip" and instead evolve to an
expanding flat de Sitter phase.

\subsubsection{The $-1/3\leq\alpha$ sub-case}\label{sec4b3}

\begin{figure}[h!]
\begin{center}
\hspace{0.4cm}\includegraphics[width=8.5cm,height=8.5cm,angle=270]{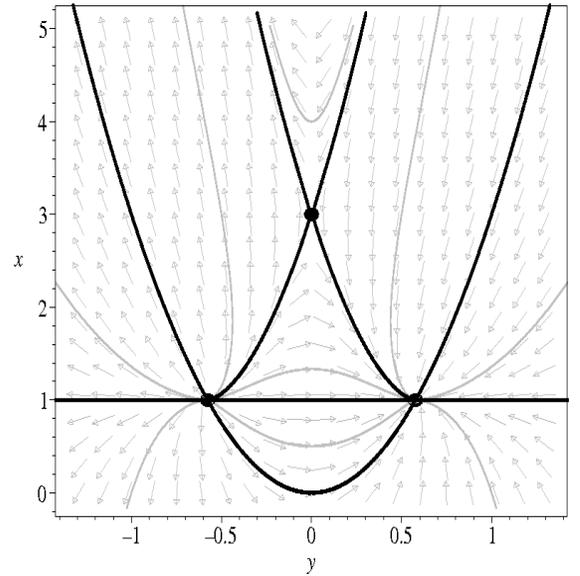}
\caption{The phase space for the low energy regime system with
$\epsilon_p=-1$ and $-1/3\leq\alpha$. The upper (lower) region
corresponds to the case {\bf E2} ({\bf F2}).}
\end{center}
\label{fig13}
\end{figure}

We now consider the system in the parameter range
$-1/3\leq\alpha$, the phase space is shown in Fig. 13. The PS
($\tilde{x}_\Lambda=(\alpha+1)^{-1}$), FFS ($y^2=x/3$) and
generalized flat de Sitter points ($dS_{\pm}$) still remain. The
flat expanding (contracting) de Sitter model is the generic future
attractor (repeller). The inner most black curve is the closed
Friedmann separatrix (CFS) and coincides with the generalized
Einstein static fixed point ($E$), which has saddle stability. The
CFS is given by:
\begin{equation}
y^2 = \frac{x}{3} - D \left[ \frac{(\alpha+1)x-1}{2}
\right]^{\frac{2}{3(\alpha+1)}}.
\end{equation}

\noindent The constant $D$ is a constant of integration and can be
fixed by the location of the fixed point $E$. The constant is
given in terms of $\alpha$ and has the form:
\begin{equation}
D=\left({3\alpha+1}\right)^{-\frac{(3\alpha+1)}{3(\alpha+1)}}.
\end{equation}

The region below the PS ($x<\tilde{x}_\Lambda$) remains
qualitatively the same. The open models in the standard matter
region ($x>\tilde{x}_\Lambda$) all evolve from a ``Big Bang" to a
expanding flat de Sitter phase. The trajectories between the FFS
and the CFS also evolve from a ``Big Bang" to a generalized
expanding flat de Sitter model with the possibility of entering a
loitering phase. The models inside the CFS can behave in one of
two ways. The trajectories above the generalized Einstein static
point represent turn-around models which evolve from a ``Big
Bang", reach a maximum $a$ (minimum $x$) and then re-collapse to a
``Big Crunch". The trajectories below evolve from a contracting de
Sitter phase to an expanding de Sitter phase and represent bounce
models.

\section{The Full system}\label{sec5}

\begin{table*}[t!]
\caption{\label{Tab7}The locations of the fixed points of the full
system. The existence conditions are also given, that is
$x,y,\in\mathbb{R}$ and $x\geq0$. To simplify the expressions
special values of $\nu$ are used which can be expressed in terms
of $\alpha$, these values are $\nu_{1}=\frac{(3\alpha+1)^2}{36} $
and $ \nu_{2}=\frac{(\alpha+1)^2}{4}$. }
\begin{ruledtabular}
\begin{tabular}{ccccc}
Name  & $x$ & $y$ & Existence ($\epsilon=+1$) & Existence ($\epsilon=-1$) \\
\hline
 & & & & \\
$M$ & $0$ & $0$ & $\nu=0$ &  $\nu=0$ \\

& & & & \\
$E_{1}$ & $ -\frac{(3\alpha+1)}{6\epsilon} + \frac{
\sqrt{(3\alpha+1)^2 - 36\epsilon\nu} }{6\epsilon}  $ & $0$ & $\nu
\leq \nu_{1}$, $\alpha < -1/3$ & $ -\nu_{1} < \nu < 0$, $\alpha > -1/3$ \\
& & & $\nu<0$, $\alpha > -1/3$ & \\

& & & & \\
$E_{2}$ & $-\frac{(3\alpha+1)}{6\epsilon} - \frac{
\sqrt{(3\alpha+1)^2 - 36\epsilon\nu} }{6\epsilon} $ & $0$ & $0 <
\nu < \nu_{1}$, $\alpha < -1/3$ &  $\nu >0$ , $\alpha < -1/3$ \\
& & & & $\nu \geq -\nu_{1}$, $\alpha > -1/3$\\

& & & & \\
$dS_{1,\pm}$ & $-\frac{(\alpha+1)}{2\epsilon} + \frac{
\sqrt{(\alpha+1)^2 - 4\epsilon\nu} }{2\epsilon}$ & $\pm
\left(-\frac{(\alpha+1)}{6\epsilon} + \frac{ \sqrt{(\alpha+1)^2 -
4\epsilon\nu} }{6\epsilon} \right)^{\frac{1}{2}}$ & $\nu \leq
\nu_{2}$, $\alpha < -1$ &  $ -\nu_{2}<\nu<0$, $\alpha > -1$ \\
& & & $\nu<0$, $\alpha > -1$ & \\

& & & & \\
$dS_{2,\pm}$ & $-\frac{(\alpha+1)}{2\epsilon} - \frac{
\sqrt{(\alpha+1)^2 - 4\epsilon\nu} }{2\epsilon}$ & $\pm \left(
-\frac{(\alpha+1)}{6\epsilon} - \frac{ \sqrt{(\alpha+1)^2 -
4\epsilon\nu} }{6\epsilon} \right)^{\frac{1}{2}}$ & $0< \nu
< \nu_{2}$, $\alpha < -1$ & $\nu>0$, $\alpha < -1$ \\
& & & & $-\nu_{2} \leq \nu$, $\alpha > -1$ \\

& & & & \\
\end{tabular}
\end{ruledtabular}
\end{table*}

\subsection{The dimensionless dynamical system}

We now consider the system of equations with the full quadratic
EoS, this can be simplified in a similar fashion to the previous
case by introducing a new set variables:
\begin{equation}
x=\frac{\rho}{|\rho_{c}|}\;,~~ y=\frac{H}{\sqrt{|\rho_{c}|}}\;,~~
\eta=\sqrt{|\rho_{c}|}t\;,~~
\nu=\frac{P_{o}}{\sqrt{|\rho_{c}|}}\;.
\end{equation}

\noindent The system of equations then become:
\begin{eqnarray}
y^{2} &=& \frac{x}{3} - \frac{K}{|\rho_{c}|a^2}, \\
y' &=& -y^{2} - \frac{1}{6} \left( 3\nu + (3\alpha +1)x + 3\epsilon x^{2} \right) ,\\
x'&=& -3 y \left( \nu + (\alpha +1)x + \epsilon x^2 \right).
\end{eqnarray}

\noindent The parameter $\epsilon$ denotes the sign of the
quadratic term, $\epsilon\in\{-1,1\}$. The parameter $\nu$ is the
normalized constant pressure term. The primes denote
differentiation with respect to the new normalized time variable
$\eta$ and only the physical region of the phase space is
considered ($x\geq0$). The fixed points and their existence
conditions are given in Table~\ref{Tab7}. The phase space
undergo's a topological change for special values of the $\nu$
parameter, these values can be expressed in terms of $\alpha$ and
are:
\begin{eqnarray}
\nu_{1} = \frac{(3\alpha+1)^2}{36},\qquad
\nu_{2}=\frac{(\alpha+1)^2}{4}.
\end{eqnarray}


\begin{center}
\begin{table}[h!]\caption{\label{Tab9} The linear stability
character of the fixed points for the full system. The stability
character is only valid for choices of parameters which are
consistent with the existence conditions and constraints given
below.}
\begin{tabular*}{0.47\textwidth}{@{\extracolsep{\fill}}ccc}
\hline \hline
Name &$\epsilon=\pm1$ & Exceptions \\
\hline
\\
$M$ & Undefined & -  \\

$E_{1}$ & Saddle  & $36\epsilon\nu\neq(3\alpha+1)^2$\\

$E_{2}$ & Center & $36\epsilon\nu\neq(3\alpha+1)^2$ \\

$\,dS_{1,+}\,$ & Attractor  & $4\epsilon\nu\neq(\alpha+1)^2$  \\

$\,dS_{1,-}\,$ & Repeller  & $4\epsilon\nu\neq(\alpha+1)^2$  \\

$\,dS_{2,+}\,$ & Saddle  & $4\epsilon\nu\neq(\alpha+1)^2$   \\

$\,dS_{2,-}\,$ & Saddle  & $4\epsilon\nu\neq(\alpha+1)^2$  \\
\\
\hline \hline
\end{tabular*}
\end{table}
\end{center}

\begin{table*}[t!]
\caption{\label{Tab8}The Eigenvalues derived from the linear
stability analysis of the fixed points for the full system. In
order to simplify the form of the eigenvalues we introduce the
following variables, $\gamma_{1}=(3\alpha+1) $, $
\gamma_{2}=(\alpha+1)$ and $\delta =
\sqrt{(\alpha+1)^2-4\epsilon\nu}$. These eigenvalues are only
valid for the choice of parameters consistent with the existence
conditions. }
\begin{ruledtabular}
\begin{tabular}{ccc}
Name  & $\lambda_{1}$ & $\lambda_{2}$  \\
\hline
 & &  \\
$M$ & $0$ & $0$  \\

& &  \\

$E_{1}$ & $ + \left( \frac{\gamma_{1}^{2}-\gamma_{1}
\sqrt{\gamma_{1}^{2}-36\epsilon\nu}-36\epsilon\nu}{18 \epsilon}
\right)^{\frac{1}{2}}$ & $ - \left(
\frac{\gamma_{1}^{2}-\gamma_{1}
\sqrt{\gamma_{1}^{2}-36\epsilon\nu}-36\epsilon\nu}{18 \epsilon}
\right)^{\frac{1}{2}}$ \\

& & \\

$E_{2}$ & $ + \left( \frac{\gamma_{1}^{2}+\gamma_{1}
\sqrt{\gamma_{1}^{2}-36\epsilon\nu}-36\epsilon\nu}{18 \epsilon}
\right)^{\frac{1}{2}}$ & $ - \left(
\frac{\gamma_{1}^{2}+\gamma_{1}
\sqrt{\gamma_{1}^{2}-36\epsilon\nu}-36\epsilon\nu}{18 \epsilon}
\right)^{\frac{1}{2}}$ \\

& &  \\

$dS_{1,\pm}$ & $ \mp \sqrt{\frac{\delta-\gamma_{2}}{6\epsilon}}
\left( 1+\frac{3\delta}{2} \right) + \left( \frac{{6\delta^2}
(3\delta-3\gamma_{2}-4)+ 8(\gamma_{2}(3\delta-1)+
\delta)}{48\epsilon}\right)^{\frac{1}{2}} $ & $ \mp
\sqrt{\frac{\delta-\gamma_{2}}{6\epsilon}} \left(
1+\frac{3\delta}{2} \right) - \left( \frac{{6\delta^2}
(3\delta-3\gamma_{2}-4)+ 8(\gamma_{2}(3\delta-1)+
\delta)}{48\epsilon}\right)^{\frac{1}{2}} $ \\

& & \\

$dS_{2,\pm}$ & $ \pm \sqrt{\frac{-(\delta+\gamma_{2})}{6\epsilon}}
\left( \frac{3\delta}{2} -1 \right) + \left( -\frac{{6\delta^2}
(3\delta+3\gamma_{2}+4)+ 8(\gamma_{2}(3\delta+1)+
\delta)}{48\epsilon}\right)^{\frac{1}{2}} $ & $ \pm
\sqrt{\frac{-(\delta+\gamma_{2})}{6\epsilon}} \left(
\frac{3\delta}{2} -1 \right) - \left( -\frac{{6\delta^2}
(3\delta+3\gamma_{2}+4)+ 8(\gamma_{2}(3\delta+1)+
\delta)}{48\epsilon}\right)^{\frac{1}{2}} $ \\

& & \\
\end{tabular}
\end{ruledtabular}
\end{table*}

\noindent As in the previous case, by existence we mean $x\geq0$
and $x,y\in\mathbb{R}$. The general eigenvalues derived from the
linear stability analysis are given in Table~\ref{Tab8}. The
linear stability character of the fixed points is given in
Table~\ref{Tab9}.

The system has six fixed points and the sign of $\epsilon$ no
longer effects the linear stability character of the fixed point
(but changes it's position in the $x-y$ plane). The first fixed
point $M$ represents a Minkowski model and is only present if
$\nu=0$, the linear stability character is undefined. The second
fixed point $E_{1}$ represents an Einstein static model and has
the linear stability character of a saddle. The third fixed point
$E_{2}$ represents an Einstein static model with the linear
stability character of a center. In general, this fixed point is
surrounded by a set of closed concentric loops representing
oscillating models. The next pair of fixed points $dS_{1,\pm}$
represents a set of generalized flat de Sitter models, the
expanding (contracting) model has attractor (repeller) stability.
The next pair of fixed points $dS_{2,\pm}$ also represents a set
of generalized flat de Sitter models, but now they have saddle
stability. The separatrix for open Friedmann models (OFS) is only
present if $\nu=0$. The parabola $y^2=x/3$ (FFS) still separates
the open and closed models. The separatrix for the closed
Friedmann models (CFS) is present for a narrow range of the
parameters and always coincides with the fixed points representing
the generalized Einstein static model, $E_{1}$. The fluid permits
two possible effective cosmological constants points, they are
given by:
\begin{eqnarray}
x_{\Lambda,1} &=&
\frac{\rho_{\Lambda,1}}{|\rho_{c}|}=-\frac{(\alpha+1)}{2\epsilon}
+ \frac{ \sqrt{\delta} }{2\epsilon} ,\\
x_{\Lambda,2} &=&
\frac{\rho_{\Lambda,2}}{|\rho_{c}|}=-\frac{(\alpha+1)}{2\epsilon}
- \frac{ \sqrt{\delta} }{2\epsilon}.
\end{eqnarray}

\noindent Where $\delta=(\alpha+1)^2-4\epsilon\nu$. There is also
a separatrix associated with each of the effective cosmological
points, which divide the regions of phantom and non-phantom
behavior. The separatrices will be referred to as the phantom
separatrix ($PS_{i}$ which corresponds to the line
$x=x_{\Lambda,i}$), with the appropriate subscript. For special
choices of parameters the separatrices coincide. The discussion of
the system will be split into the two categories, $\epsilon=+1$
and $\epsilon=-1$.

\subsection{The $\epsilon=+1$ case}\label{sec5a}

We first consider the system when we have a positive quadratic
energy density term ($\epsilon=+1$). The dynamical system can be
further sub-divided into sub-cases with different values of the
parameters $\alpha$ and $\nu$. The various subcases have been
highlighted in Table~\ref{Tab10}.
\begin{center}
\begin{table}[h!]\caption{\label{Tab10} The various sub-cases of
the full system when $\epsilon=+1$. The figure numbers given in
bold, indicate the choice of variables for which the state space
is qualitatively different to previously discussed cases.}
\begin{tabular*}{0.47\textwidth}{@{\extracolsep{\fill}}cccc}
\hline \hline
                        & $\alpha<-1$   & $-1\leq\alpha<-1/3$   & $-1/3\leq\alpha$  \\
\hline
& & & \\
$\nu>\nu_{1}$           &    FIG.10     &      FIG.10           &      FIG.10       \\
$\nu=\nu_{1}$           &  {\bf FIG.14} &   {\bf FIG.14}        &      FIG.10       \\
$\nu_{2}<\nu<\nu_{1}$   &  {\bf FIG.15} &   {\bf FIG.15}        &      FIG.10       \\
$\nu=\nu_{2}$           &  {\bf FIG.16} &   {\bf FIG.15}        &      FIG.10       \\
$0<\nu<\nu_{2}$         &  {\bf FIG.17} &   {\bf FIG.15}        &      FIG.10       \\
$\nu=0$                 &    FIG.1      &      FIG.3            &      FIG.4        \\
$\nu<0$                 &    FIG.13     &      FIG.13           &      FIG.13       \\
& & & \\
\hline \hline
\end{tabular*}
\end{table}
\end{center}

The majority of sub-cases result in a phase space diagram which is
qualitatively similar to cases discussed in previous sections.
That is the qualitative behavior of trajectories is the same even
though  the functional form of $\rho(a)$ is different. The figure
numbers not in bold (standard text) indicate choices of variable
for which the phase space is qualitatively similar to a previous
case, with the following differences:
\begin{itemize}
\item The regions which corresponded to different types of
behavior of the fluid now change (replaced by new $\rho(a)$
behavior):
\begin{itemize}
\item The case {\bf D1} $\to$ {\bf G1},
\item The case {\bf E2} $\to$ {\bf I3},
\item The case {\bf F2} $\to$ {\bf I2},
\end{itemize}
\item The Type I singularities are now replaced by Type III
singularities.
\end{itemize}

There is a narrow range of the parameters for which the state
space is qualitatively different. The figure numbers given in bold
in Table~\ref{Tab10}, indicate the choice of variables for which
the state space is qualitatively different to previously discussed
cases. We will now discuss the four sub-cases which are different
to those discussed in previous sections.

\subsubsection{The $\alpha<-1$,  $\nu=\nu_{1}$ sub-case}\label{sec5a1}

\begin{figure}[h!]
\begin{center}
\hspace{0.4cm}\includegraphics[width=8.5cm,height=8.5cm,angle=270]{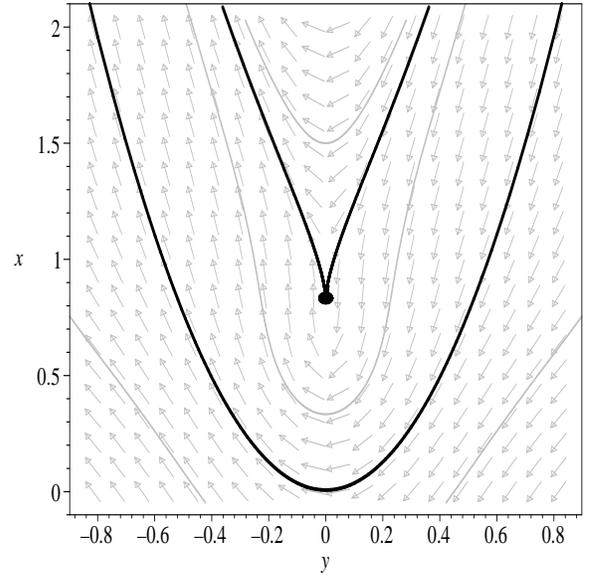}
\caption{The phase space for the full system with $\epsilon=+1$,
$\alpha < -1$ and $\nu=\nu_{1}$ (additionally when $\alpha < -1/3$
and  $\nu=\nu_{1}$). The entire region corresponds to the case
{\bf G1}.}
\end{center}
\label{fig14}
\end{figure}

The phase space of the system when $\alpha<-1$ and  $\nu=\nu_{1}$
is shown in Fig. 14. As before the black lines represent
separatrix, grey lines represent example trajectories and fixed
points are represented by black dots. The fluid behaves in a
standard manner and corresponds to the case {\bf G1}. This choice
of parameters results in the two Einstein points ($E_{i}$)
coinciding. The resulting fixed point is highly non-linear and
cannot be classified into the standard linear stability categories
as in previous cases. The fixed point coincides with the CFS and
the parabola is the FFS. The open models are all non-physical as
they evolve to the $x<0$ region of the phase space. The models
between the FFS and the CFS represent turn-around models which
evolve from a Type III singularity\footnote{As with the case of
the high energy EoS, the Type III singularity is a generic feature
of the fully quadratic EoS.}, evolve to a maximum $a$ (minimum
$x$) and then re-collapse. The trajectories above the CFS also
represent similar turn-around models.

\subsubsection{The $\alpha<-1$, $\nu_{2}<\nu<\nu_{1}$ sub-case}\label{sec5a2}

\begin{figure}[h!]
\begin{center}
\hspace{0.4cm}\includegraphics[width=8.5cm,height=8.5cm,angle=270]{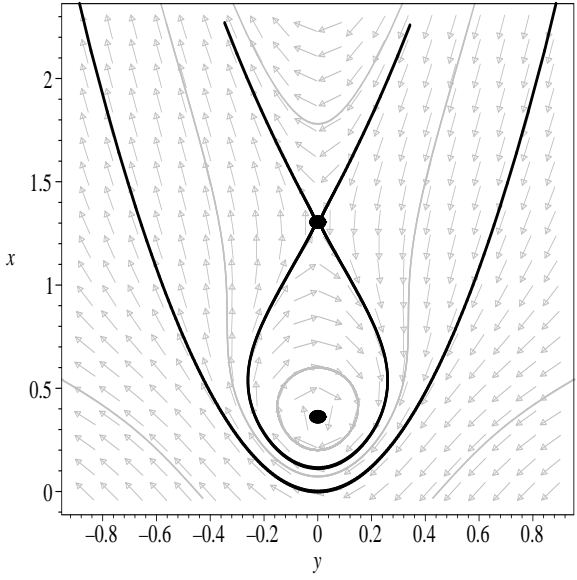}
\caption{The phase space for the full system with $\epsilon=+1$,
$\alpha < -1$ and $\nu_{2} < \nu < \nu_{1}$ (additionally when $-1
< \alpha < -1/3$ and  $0 < \nu < \nu_{1}$). The entire region
corresponds to the case {\bf G1}.}
\end{center}
\label{fig15}
\end{figure}

The phase space of the system when $\alpha<-1$ and $\nu_{2} < \nu
< \nu_{1}$ is shown in Fig. 15. The fluid behaves in a standard
manner and corresponds to the case {\bf G1}. The Einstein fixed
point of the previous case splits into two individual Einstein
fixed points ($E_{i}$) via bifurcation. The first Einstein fixed
point ($E_{1}$) coincides with the CFS, while the second Einstein
fixed point ($E_{2}$) is located inside the lower region enclosed
by the CFS. Only the trajectories above the FFS differ from the
previous case. The trajectories between the CFS and FFS still
represent turn-around models which evolve from a Type III
singularity but may now enter a loitering phase. The trajectories
inside the CFS and above the Einstein fixed point ($E_{1}$),
evolve from a Type III singularity, reach a maximum $a$ and then
re-collapse to a Type III singularity. The trajectories below
$E_{1}$ represent closed oscillating models, the closed loops are
centered on the second Einstein fixed point ($E_{2}$).

\subsubsection{The $\alpha<-1$, $\nu=\nu_{2}$ sub-case}\label{sec5a3}

\begin{figure}[h!]
\begin{center}
\hspace{0.4cm}\includegraphics[width=8.5cm,height=8.5cm,angle=270]{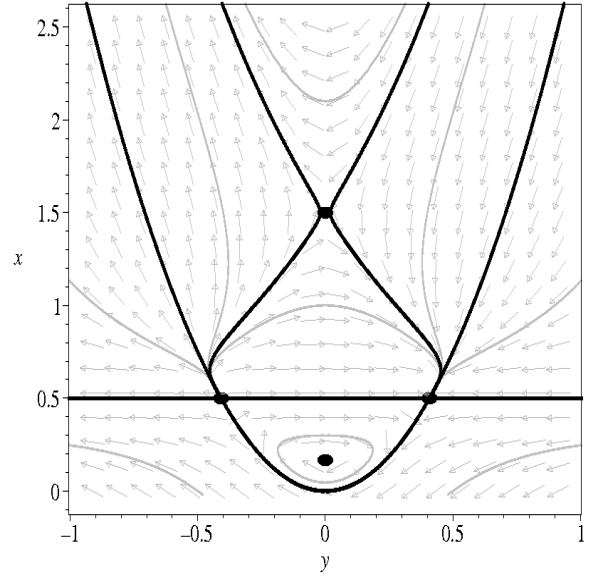}
\caption{The phase space for the full system with $\epsilon=+1$,
$\alpha < -1$ and $\nu=\nu_{2}$. The upper (lower) region
corresponds to the case {\bf H2} ({\bf H1}).}
\end{center}
\label{fig16}
\end{figure}

The phase space of the system when $\alpha<-1$ and $ \nu =
\nu_{2}$ is shown in Fig. 16. The fluid behaves in a standard
manner in both regions and the upper (lower) region corresponds to
the case {\bf H2} ({\bf H1}). This choice of parameters results in
the two sets of generalized de Sitter points ($dS_{i,\pm}$)
coinciding. The fixed points coincide with $PS_{i}$
($x=x_{\Lambda,i}$) and the FFS. The resulting fixed points are
highly non-linear, the points have shunt stability along the FFS
direction and the generalized expanding (contracting) de Sitter
point has attractor (repeller) stability along the $PS_{i}$
direction. The two Einstein points ($E_{i}$) and the CFS are still
present. In the  $x<x_{\Lambda,i}$ region, the open models are all
non-physical as they evolve to the $x<0$ region and the closed
models represent oscillating models which are centered on the
Einstein point ($E_{2}$) with center linear stability. In the
$x>x_{\Lambda,i}$ region, the open models are asymptotic to a Type
III singularity in the past and a expanding flat de Sitter phase
($dS_{i,+}$) in the future. The trajectories between the FFS and
the CFS evolve from a Type III singularity to $dS_{i,+}$ with the
possibility of entering a loitering phase. The models inside the
CFS and above the $E_{1}$ point represent turn-around models which
asymptotically approach a Type III singularity. The closed models
inside the CFS and below the $E_{2}$ point are asymptotic to a
contracting de Sitter model phase ($dS_{i,-}$) in the past and a
expanding de Sitter phase ($dS_{i,+}$) in the future. The generic
attractor in the $x>x_{\Lambda,i}$ region is the $dS_{i,+}$ fixed
point.

\subsubsection{The $\alpha<-1$, $0<\nu<\nu_{2}$ sub-case}\label{sec5a4}

\begin{figure}[h!]
\begin{center}
\hspace{0.4cm}\includegraphics[width=8.5cm,height=8.5cm,angle=270]{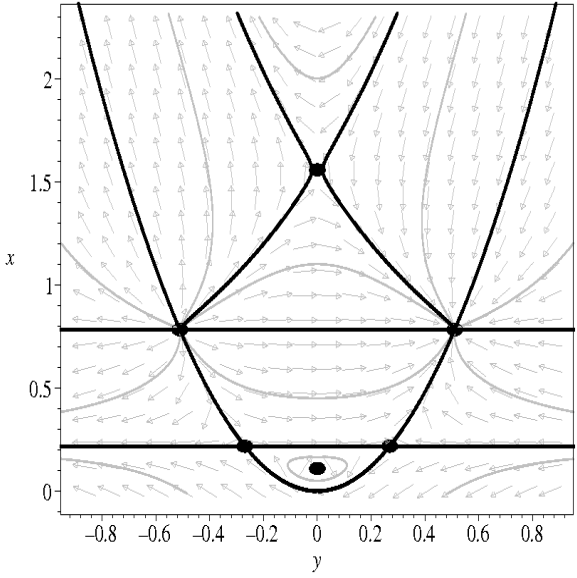}
\caption{The phase space for the full system with $\epsilon=+1$,
$\alpha < -1$ and $0<\nu<\nu_{2}$. The upper, middle and lower
regions correspond to the cases {\bf I3}, {\bf I2} and {\bf I1}
respectively.}
\end{center}
\label{fig17}
\end{figure}

The phase space of the system when $\alpha<-1$ and $0<\nu<\nu_{2}$
is shown in Fig. 17. The upper (lower) horizontal line is the
$PS_{1}$ ($PS_{2}$). The region above $PS_{1}$ corresponds to the
case {\bf I3} and is qualitatively similar to the {\bf H2} region
in the previous sub-case. The region below $PS_{2}$ corresponds to
the case {\bf I1} and is qualitatively similar to the {\bf H1}
region in the previous sub-case. The set of generalized flat de
Sitter fixed points ($dS_{i,\pm}$) of the previous case split into
two sets of generalized flat de Sitter fixed points via
bifurcation. The upper (lower) set of generalized de Sitter
points, $dS_{1,\pm}$ ($dS_{2,\pm}$) have attractor/repeller
(saddle) stability. The region between $PS_{1}$ and $PS_{2}$
corresponds to the case {\bf I2} and the fluid behaves in a
phantom manner. The open models in this region are asymptotic to
open de Sitter models in the past and flat de Sitter models in the
future. The closed models in the phantom region represent phantom
bounce models which asymptotically approach a expanding
(contracting) de Sitter phases in the future (past).

\subsection{The $\epsilon=-1$ case}\label{sec5b}

We now consider the system when we have a negative quadratic
energy density term ($\epsilon=-1$). As before the system can be
sub-divided into various sub-cases with different values of
parameters of $\alpha$ and $\nu$. The various sub-cases have been
highlighted in Table~\ref{Tab11}
\begin{center}
\begin{table}[h!]\caption{\label{Tab11}  The various sub-cases of
the $\epsilon=-1$ full system. The figure numbers given in bold,
indicate the choice of variables for which the phase space is
qualitatively different to previous cases.}
\begin{tabular*}{0.47\textwidth}{@{\extracolsep{\fill}}cccc}
\hline \hline
                        & $\alpha<-1$   & $-1\leq\alpha<-1/3$   & $-1/3\leq\alpha$  \\
\hline
& & & \\
$\nu>0$                 &    FIG.8      &      FIG.8            &      FIG.8        \\
$\nu=0$                 &    FIG.5      &      FIG.6            &      FIG.7        \\
$-\nu_{1}<\nu<0$        &    FIG.11     &   {\bf FIG.20}        &    {\bf FIG.18}   \\
$\nu=-\nu_{1}$           &    FIG.11     &   {\bf FIG.20}        &    {\bf FIG.19}   \\
$-\nu_{2}<\nu<-\nu_{1}$  &    FIG.11     &   {\bf FIG.20}        &    {\bf FIG.20}   \\
$\nu=-\nu_{2}$           &    FIG.11     &   {\bf FIG.21}        &    {\bf FIG.21}   \\
$\nu<-\nu_{2}$           &    FIG.11     &      FIG.11           &      FIG.11       \\
& & & \\
\hline \hline
\end{tabular*}
\end{table}
\end{center}

\begin{figure}[h!]
\begin{center}
\hspace{0.4cm}\includegraphics[width=8.5cm,height=8.5cm,angle=270]{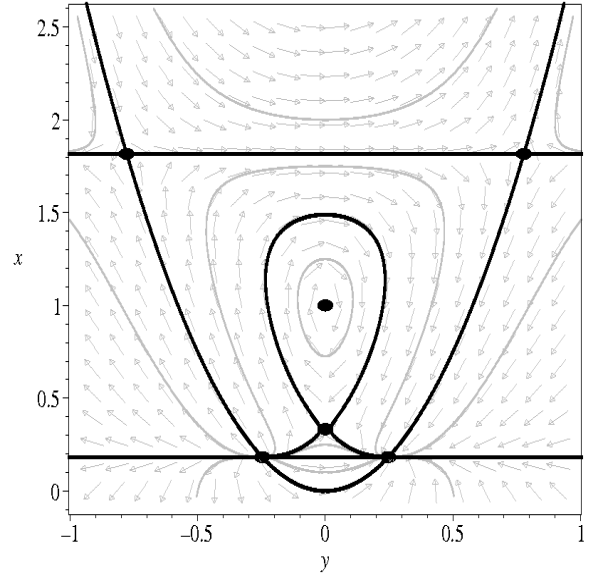}
\caption{The phase space for the full system with $\epsilon=-1$,
$\alpha > -1/3$ and $-\nu_{1} < \nu < 0$. The upper, middle and
lower regions correspond to the cases {\bf I6}, {\bf I5} and {\bf
I4} respectively.}
\end{center}
\label{fig18}
\end{figure}

\noindent As before, the figure numbers not in bold (standard
text) indicate choices of variable for which the phase space is
qualitatively similar to a previous case, with the following
differences:
\begin{itemize}
\item The regions which corresponded to different types of
behavior of the fluid now change (replaced by new form of
$\rho(a)$):
\begin{itemize}
\item The case {\bf D2} $\to$ {\bf G2},
\item The case {\bf E1} $\to$ {\bf I6},
\item The case {\bf F1} $\to$ {\bf I5},
\end{itemize}
\item The Type I singularities are now replaced by Type III
singularities.
\end{itemize}

\noindent There are choices of parameters for which the phase
space is different (figure numbers in bold in Table~\ref{Tab11})
and these four sub-cases will be discussed in the following
sections.
\begin{figure}[t!]
\begin{center}
\hspace{0.4cm}\includegraphics[width=8.5cm,height=8.5cm,angle=270]{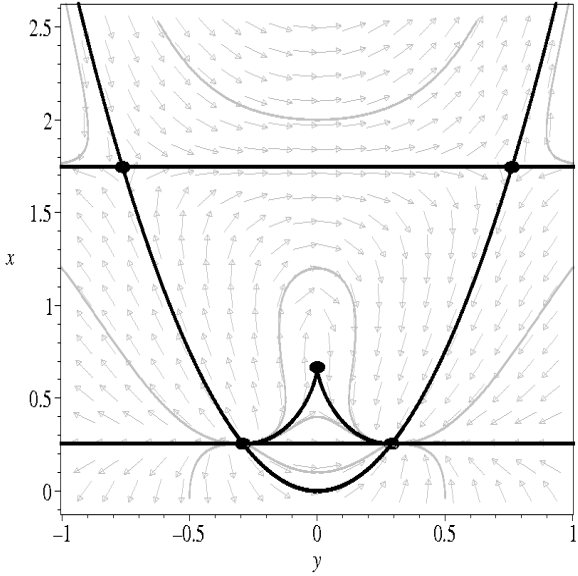}
\caption{The phase space for the full system with $\epsilon=-1$,
$\alpha > -1/3$ and $\nu = -\nu_{1}$. The upper, middle and lower
regions correspond to the cases {\bf I6}, {\bf I5} and {\bf I4}
respectively.}
\end{center}
\label{fig19}
\end{figure}

\subsubsection{The $\alpha>-1/3$, $-\nu_{1}<\nu<0$ sub-case}\label{sec5b1}

The phase space of the system when $\alpha>-1/3$ and
$-\nu_{1}<\nu<0$ is shown in Fig. 18. The upper (lower) horizontal
line at $x=x_{\Lambda,2}$ ($x=x_{\Lambda,1}$) is the $PS_{2}$
($PS_{1}$) (they have swapped position with respect to the
$\epsilon=+1$ case). The region above $PS_{2}$ corresponds to the
case {\bf I6}, the region below $PS_{1}$ corresponds to the case
{\bf I4} and the fluid behaves in a phantom manner in both
regions. The region between $PS_{1}$ and $PS_{2}$ corresponds to
the case {\bf I5} and the fluid behaves in a standard manner. The
lower set of generalized de Sitter points ($dS_{1,\pm}$ - at the
intersection of $PS_{1}$ and FFS) have attractor/repeller
stability, while the upper set ($dS_{2,\pm}$ - at the intersection
of $PS_{2}$ and FFS) have saddle stability. The CFS is located in
between $PS_{1}$ and $PS_{2}$ and coincides with the Einstein
point ($E_{1}$). The open models in the $x< x_{\Lambda,1}$ region
(the case {\bf I4}) are all non-physical as they evolve from the
$x<0$ region of the phase space. The closed models in this region
represent phantom bounce models which evolve from a contracting de
Sitter phase ($dS_{1,-}$) to a expanding de Sitter phase
($dS_{1,+}$). The open models in the standard region
($x_{\Lambda,1}< x< x_{\Lambda,2}$ corresponding to the case {\bf
I5} ) are asymptotic to a generalized open de Sitter model in the
past and generalized flat de Sitter model in the future (the
future attractor has lower $x$ and $y$). The models between the
CFS and the FFS in this region represent bounce models which
evolve from a contracting de Sitter phase to a expanding de Sitter
phase with the possibility of entering a loitering phase. The
models enclosed by the CFS can be split into two groups. The
models above the fixed point, $E_{1}$ represent oscillating
models, the closed loops are centered on the fixed point $E_{2}$.
The models below the fixed point, $E_{1}$ represent bounce models
which evolve from $dS_{1,-}$ to $dS_{1,+}$. In the
$x>x_{\Lambda,2}$ region (the case {\bf I6}) the open models are
asymptotic to generalized open de Sitter models in the past and a
Type III singularity in the future. The closed models in this
region represent phantom bounce models which evolve from a Type
III singularity, reach a minimum $a$ (minimum $x$) and then evolve
to a Type III singularity. The generalized expanding flat de
Sitter model, $dS_{1,+}$ (Type III singularity) is the generic
future attractor in the region $x<x_{\Lambda,2}$
($x>x_{\Lambda,2}$). The trajectories in the regions,
$x<x_{\Lambda,1}$ and $x>x_{\Lambda,2}$ remain qualitatively
similar in the following two cases (Fig.19,~20).

\subsubsection{The $\alpha>-1/3$, $\nu=-\nu_{1}$ sub-case}\label{sec5b2}

The phase space of the system when $\alpha>-1/3$ and
$\nu=-\nu_{1}$ is shown in Fig. 19. The phase space is equivalent
to the previous sub-case, except for the region $x_{\Lambda,1}< x<
x_{\Lambda,2}$ (the case {\bf I5}). The open models in this region
are still asymptotic to generalized open (flat) de Sitter models
in the past (future). The behavior of the closed models has now
changed, there are no longer trajectories representing oscillating
models. The two generalized Einstein fixed points ($E_{i}$) have
now coalesced to form one fixed point via bifurcation. The closed
models above $E_{i}$ represent bounce models which evolve from
$dS_{1,-}$ to $dS_{1,+}$, with the possibility of entering a
loitering phase. The closed models below $E_{i}$ represent bounce
models which evolve from $dS_{1,-}$ to $dS_{1,+}$ without entering
a loitering phase.

\subsubsection{The $\alpha>-1/3$, $-\nu_{2}<\nu<-\nu_{1}$ sub-case}\label{sec5b3}

\begin{figure}[t!]
\begin{center}
\hspace{0.4cm}\includegraphics[width=8.5cm,height=8.5cm,angle=270]{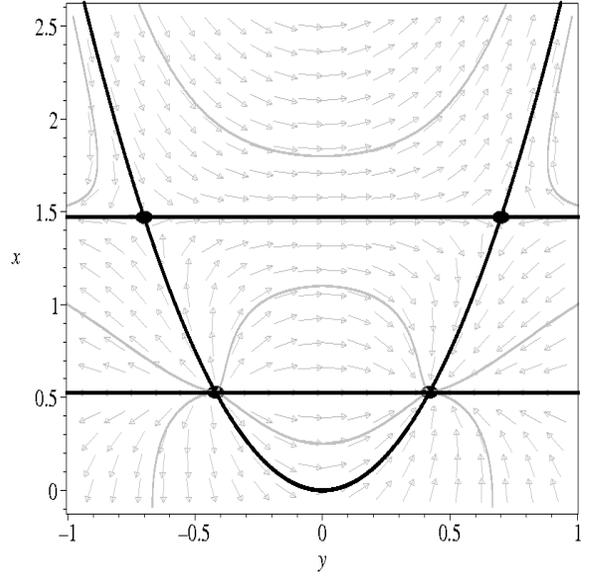}
\caption{The phase space for the full system with $\epsilon=-1$,
$\alpha > -1/3$ and $-\nu_{2} < \nu < -\nu_{1}$ (additionally when
$-1 < \alpha < -1/3$ and  $-\nu_{2} < \nu < 0$). The upper, middle
and lower regions correspond to the cases {\bf I6}, {\bf I5} and
{\bf I4} respectively.}
\end{center}
\label{fig20}
\end{figure}

The phase space of the system when $\alpha>-1/3$ and
$-\nu_{2}<\nu<-\nu_{1}$ is shown in Fig. 20. The phase space is
qualitatively similar to the previous sub-cases except for the
$x_{\Lambda,1}< x< x_{\Lambda,2}$ region. There are no longer any
fixed points representing generalized Einstein static models and
the CFS is no longer present. The open models in the region behave
as in previous sub-cases. The closed models in the region
represent bounce model, which evolve to a expanding (collapsing)
de Sitter phase in the future (past) without the possibility of
entering a loitering phase.

\subsubsection{The $\alpha>-1/3$, $\nu=-\nu_{2}$ sub-case}\label{sec5b4}

\begin{figure}[t!]
\begin{center}
\hspace{0.4cm}\includegraphics[width=8.5cm,height=8.5cm,angle=270]{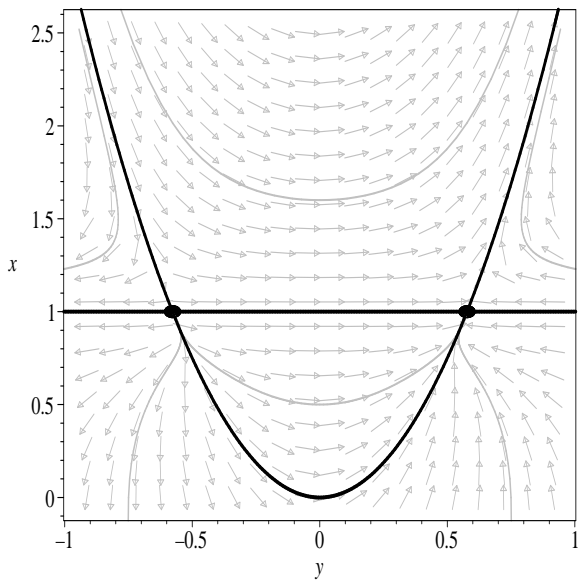}
\caption{The phase space for the full system with $\epsilon=-1$,
$\alpha > -1/3$ and $\nu = -\nu_{2}$ (additionally when $-1 <
\alpha < -1/3$ and  $\nu = -\nu_{2}$). The upper (lower) region
corresponds to the case {\bf H4} ({\bf H3}).}
\end{center}
\label{fig21}
\end{figure}

The next case is the phase space of the system when $\alpha>-1/3$
and $\nu=-\nu_{2}$ and is shown in Fig. 21. The fluid behaves in a
phantom manner in both regions and the upper (lower) region
corresponds to the case {\bf H4} ({\bf H3}). The two sets of
generalized de Sitter points ($dS_{i,\pm}$) have now coalesced
into a single set of generalized de Sitter points ($dS_{\pm}$)
which are located at the intersection of the FFS and the $PS_{i}$
which have also coalesced to form a single separatrix
($x_{\Lambda,1} = x_{\Lambda,2}$). The resulting fixed points are
highly non-linear, the points have shunt stability along the FFS
direction and the generalized expanding (contracting) de Sitter
point has attractor (repeller) stability along the $PS_{i}$
direction. The Type III singularity is the generic attractor in
the upper region ($x>x_{\Lambda,i}$) and the $dS_{i,+}$ is the
generic attractor in the lower region ($x<x_{\Lambda,i}$).

\section{Discussion and Conclusions}\label{sec6}

In this paper we have systematically studied the dynamics of
homogeneous and isotropic cosmological models containing a fluid
with a quadratic EoS. This has it's own specific interest (see
Section I for a variety of motivations) and serves as a simple
example of more general EoS's. It can also be taken to represent
the truncated Taylor expansion of any barotropic EoS, and as such
it serves (with the right choice of parameters) as a useful
phenomenological model for dark energy, or even UDM. Indeed, we
have shown the dynamics to be very different and much richer than
the standard linear EoS case, finding  that an almost generic
feature of the evolution is the existence of an accelerated phase,
most often asymptotically de Sitter,  thanks to the appearance of
an {\it effective cosmological constant}. Of course to properly
build physical cosmological models would require to consider the
quadratic EoS for dark energy or UDM together with standard matter
and radiation. Our analysis was aimed instead to derive and
classify the large variety of different dynamical effects that the
quadratic EoS fluid has when is the dominant component. In this
respect, it should be noticed that a positive quadratic term in
the EoS allows, in presence of another fluid such as radiation,
equi-density between the two fluid to occur twice, i.e. the
quadratic EoS fluid can be dominant at early and late times, and
subdominant in an intermediate era.

In Section II we have made some general remarks, mostly based on
conservation of energy only and as such valid independently of any
specific theory of gravity. We have also given the various
possible functional forms of the energy density as a function of
the scale factor, $\rho(a)$, and listed the many subcases, grouped
in three main cases, what we call: {\it i)}  the high energy
models (no constant $P_o$ term); {\it ii)} the low energy  affine
EoS with no quadratic term; {\it iii)} the complete quadratic EoS.

The quadratic term in the EoS affects the high energy behavior as
expected but can additionally affect the dynamics at relatively
low energies. First, in Section III, we have concentrated on the
high energy models. The specific choice of parameters fixes the
behavior of the fluid, it can behave in a phantom or standard
manner. In the case of phantom behavior, $\rho$ can tend to zero
at early times and either tend to an effective cosmological
constant ({\bf C1}) or a Type III singularity ({\bf A2}) at late
times. Alternatively $\rho$ can also tend to an effective
cosmological constant in the past ({\bf B2}) and a Type III
singularity at late times. When the fluid behaves in a standard
manner, it can tend to a Type III singularity at early times, with
$\rho$ either tending to zero ({\bf A1}) or to an effective
cosmological constant ({\bf B1}) at late times. The fluid can also
behave as an effective cosmological constant at early times with
$\rho$ decaying away at late times ({\bf C2}). The effective
cosmological constant allows for the existence of  generalized
Einstein static ($E$) and flat de Sitter fixed ($dS_{\pm}$) points
which modify the late time behavior. The main new feature is the
existence of models which evolve from a Type III singularity and
asymptotically approach a flat de Sitter model ($dS_{+}$).  Of
specific interest are the closed models of this type, which can
also evolve through an intermediate  loitering phase.

Neglecting the quadratic term, in Section IV we have considered
the low energy models with affine EoS. As expected, the constant
term in the quadratic EoS affects the relatively low energy
behavior. It can result in a variety of qualitatively different
dynamics with respect to those of the linear EoS case. Again, the
fluid can have a phantom or standard behavior. When the fluid
behaves in a phantom manner, $\rho$ can tend to an effective
cosmological constant ({\bf F2}), or  can tend to a Type I (``Big
Rip") singularity ({\bf D2}) at late times. Alternatively, $\rho$
can also tend to an effective cosmological constant in the past
and a Big Rip in the future({\bf E1}). When the fluid behaves in a
standard manner, we recover the linear EoS at early times and
$\rho$ can either tend to zero ({\bf D1}) or to an effective
cosmological constant ({\bf E2}) at late times. The fluid can also
behave as an effective cosmological constant at early times, with
$\rho$ decaying away at late times ({\bf F1}). The effective
cosmological constant allows for the existence of new fixed
points($E$ and $dS_{\pm}$). Comparing with standard linear EoS
cosmology, the most interesting differences are new closed models
which oscillate indefinitely and new closed models which exhibit
phantom behavior which do not terminate in a ``Big Rip", but
asymptotically approach an expanding flat de Sitter model (flat
and closed models where the fluid behaves as case {\bf F2}).

When we study the dynamics of the system with the complete
quadratic EoS, Section V, we see the appearance of new fixed
points representing generalized Einstein and de Sitter models
which are not present in the high/low energy systems. The various
models of the simplified systems are present in the full system
(but with differing $\rho(a)$), but there are also models with
qualitatively new behavior. As with the previous cases, in the
case of phantom behavior, $\rho$ can tend to zero at early times
and either tend to an effective cosmological constant ({\bf H3}
and {\bf I4}) or a Type III singularity ({\bf G2}) at late times.
Alternatively $\rho$ can also tend to an effective cosmological
constant in the past ({\bf H4} and {\bf I6}) and a Type III
singularity at late times. Finally, in the phantom case $\rho$ can
also tend to an effective cosmological constant both in the past
and future ({\bf I2}). In the case of standard behavior the fluid
can tend to a Type III singularity at early times, with $\rho$
either tending to zero ({\bf G1}) or to an effective cosmological
constant ({\bf H2} and {\bf I3}) at late times. The fluid can also
behave as an effective cosmological constant at early times with
$\rho$ decaying away at late times ({\bf H1} and {\bf I1}).
Finally, in the standard fluid case $\rho$ can also tend to an
effective cosmological constant both in the past and future ({\bf
I5}). There are  models which evolve from a Type III singularity,
reach a maximum $a$ (minimum $x$) and then evolve to Type III
singularity. These also enter a loitering phase before and after
the turn around point. We also see bounce models which enter a
loitering phase and asymptotically tend to generalized expanding
(contracting) de Sitter models at late (early) times.

Of specific interest are models which evolve from a Type III
singularity as opposed to the standard ``Big Bang" ({\bf A1, B1}).
The simplest models of this type correspond to the high energy EoS
with a positive quadratic term (is possible to recover standard
behavior at late times). For these models the positive quadratic
energy density term has the potential to force the initial
singularity to be isotropic. The effects of such a fluid on
anisotropic Bianchi I and V models is investigated in Paper
II~\cite{AB}. This is  achieved by carrying out a dynamical
systems analysis of these models. Additionally, using a linearized
perturbative treatment we study the behavior of inhomogeneous and
anisotropic perturbations at the singularity. The singularity is
itself represented by an isotropic model and, If the perturbations
of the latter decay in the past,  this model represents the  local
past attractor in the larger phase space of inhomogeneous and
anisotropic models (within the validity of the  perturbative
treatment). This would mean that  in  inhomogeneous anisotropic
models  with a  positive non-linear term (at least quadratic)   in
the EoS isotropy is   a natural outcome of {\it generic initial
conditions}, unlike in the standard linear EoS case where generic
cosmological models are, in GR, highly anisotropic in the past.

\acknowledgments KNA is supported by PPARC (UK). MB is partly
supported by a visiting grant by MIUR (Italy). The authors would
like to thank Chris Clarkson, Mariam Bouhmadi-L\'{o}pez and Roy
Maartens for useful comments and discussions.


\end{document}